\documentclass[secnumarabic,aps,pra,amsfonts,twocolumn,%
showkeys,floatfix%
]{revtex4-1}

\usepackage{bm,amsfonts}
\usepackage{sidecap}
\usepackage{placeins}
\usepackage{graphicx}
\newcommand{\gl}[1]{(\ref{#1})}

\newtheorem{thm}{Theorem}[section]

\begin{document}
\title{The structural distortion in 
       antiferromagnetic LaFeAsO investigated by a group-theoretical approach}  
\author{Ekkehard Kr\"uger}
\author{Horst P. Strunk}
\affiliation{Institut f\"ur Materialwissenschaft, Materialphysik,
  Universit\"at Stuttgart, D-70569 Stuttgart, Germany}
%
\date{\today}
\begin{abstract}
  As experimentally well established, undoped LaFeAsO is antiferromagnetic
  below 137K with the magnetic moments lying on the Fe sites. We determine the
  orthorhombic body-centered group $Imma$ (74) as the space group of the
  experimentally observed magnetic structure in the undistorted lattice, i.e.,
  in a lattice possessing no structural distortions in addition to the
  magnetostriction.  We show that LaFeAsO possesses a partly filled ``magnetic
  band'' with Bloch functions that can be unitarily transformed into optimally
  localized Wannier functions adapted to the space group $Imma$. This finding
  is interpreted in the framework of a nonadiabatic extension of the
  Heisenberg model of magnetism, the nonadiabatic Heisenberg model. Within
  this model, however, the magnetic structure with the space group $Imma$ is
  not stable but can be stabilized by a (slight) distortion of the crystal
  turning the space group $Imma$ into the space group $Pnn2$ (34).  This
  group-theoretical result is in accordance with the experimentally observed
  displacements of the Fe and O atoms in LaFeAsO as reported by Clarina de la
  Cruz et al. [nature 453, 899 (2008)].
\end{abstract}
\maketitle

\section{Introduction}
Undoped LaFeAsO undergoes an abrupt structural distortion from tetragonal to
orthorhombic \cite{nomura,kitao,nakai} or to monoclinic \cite{clarina}
symmetry at $\sim\! 155K$ as well as an antiferromagnetic spin ordering
transition at $\sim\!  137K$~\cite{clarina, nomura,kitao,nakai}. Clarina de la
Cruz et al.~\cite{clarina} studied with high accuracy the structural
distortion of LaFeAsO at $4 K$ by neutron-scattering experiments and found the
Fe and O atoms to be slightly shifted from their positions in the tetragonal
phase. In agreement with the space group $P4/nmm$, the $z$ coordinates of iron
and oxygen are {\em exactly} $z = 1/2$ and $z = 0$, respectively, in the
tetragonal phase. The values $\Delta z$ of the displacements at $4 K$ in $\pm
z$ direction are reported to $\Delta z = 0.0006$ and $\Delta z = 0.0057$ for
the iron and oxygen atoms, respectively~\cite{clarina}. These displacements
cannot be simply explained by magnetostriction.

In the framework of a nonadiabatic extension of the Heisenberg
model~\cite{hei}, called nonadiabatic Heisenberg model
(NHM)~\cite{enhm,ybacuo6, josybacuo7}, the magnetic order and the
low-temperature structural distortion in LaFeAsO may be understood by
group-theoretical methods. While the non-magnetic distortion of LaFeAsO
between $155 K$ and $137 K$ shall be discussed in a following paper in the
context of the superconducting state in the doped material, the present paper
investigates the magnetic structure and the associated structural distortions
below $137 K$.


 \begin{figure*}[t]
  \includegraphics[width=.65\textwidth,angle=-90]{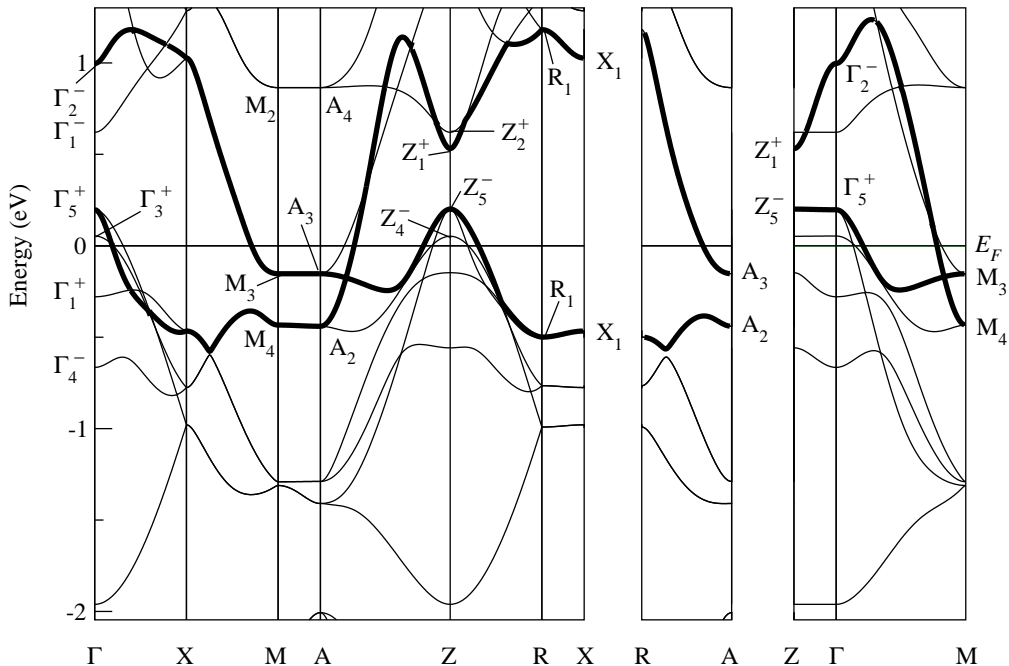}
  \caption{ Band structure of tetragonal LaFeAsO as calculated by the FHI-aims
    program~\cite{blum1,blum2}, with symmetry labels determined by the
    authors. The symmetry labels can be identified from
    Table~\ref{tab:rep129}, see Appendix. The bold line shows the magnetic
    band (as defined in Sec.~\ref{sec:bf}) consisting of two branches.
  \label{fig:bandstr}
}
 \end{figure*}


 In Sec.~\ref{sec:undistorted} we shall determine the space group $S$ of the
 experimentally observed magnetic structure in LaFeAsO under the assumption
 that there are no distortions in addition to the magnetostriction. We shall
 show that $S$ is compatible with the symmetry of the Bloch functions of the
 energy band of LaFeAsO denoted in Fig.~\ref{fig:bandstr} by the bold line. In
 the following Sec.~\ref{sec:stability} we will ask whether $S$ may be the
 space group of a {\em stable} magnetic structure, and in
 Sec.~\ref{sec:subgroups} we shall present the concept of ``allowed space
 groups''. Finally, in Sec.~\ref{sec:distortion} we shall propose that the
 experimentally observed distortions of the crystal correspond to a change
 from the non-allowed space group $S$ of the undistorted crystal to an allowed
 space group.

The NHM does not distinguished between orbital and spin moments.  Therefore,
we always speak of ``magnetic moments'' which may consist of both orbital and
spin moments.

\section{Coordinate systems}
\label{sec:coordinates}
\begin{figure}[t]
   \includegraphics[width=.4\textwidth,angle=0]{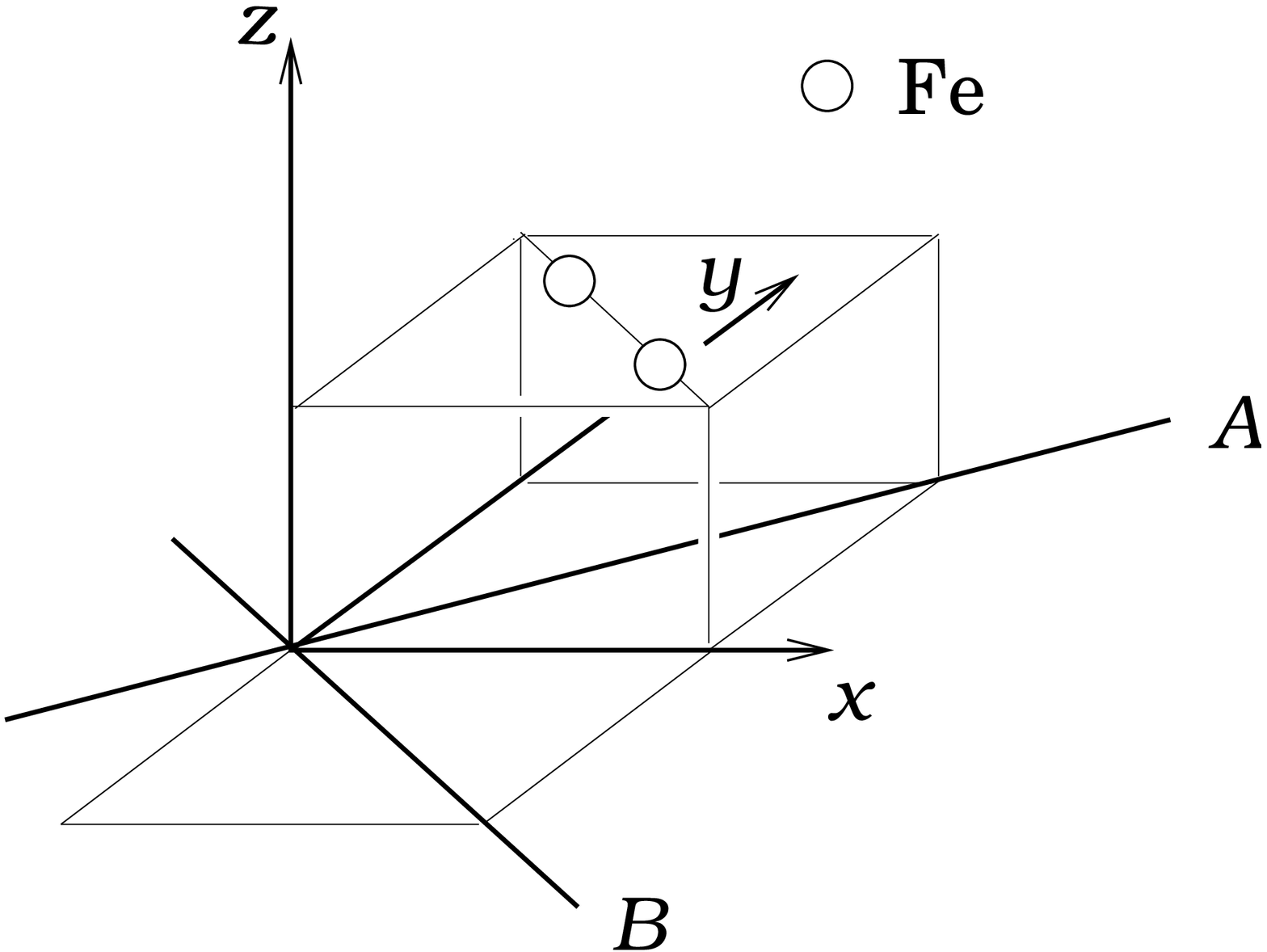}
   \caption{Cartesian coordinate system with the $A$ and $B$ axes defining
     the point group operations as used in this paper. The $x, y, z$
     coordinate system is identical with the $x, y, z$
     coordinate systems in Fig.~\ref{fig:structures}. The $A$ and $B$ axes 
     lie in the plane spanned by the $x$ and $y$ axes and, in all the 
     structures in 
     Figs.~\ref{fig:structures} (a), (b), and (c), they have the direction of a 
     lattice translation. Hence, in all the structures they are perpendicular
     to one another, while the $x$ axis is  perpendicular to the $y$ axis only
     in the tetragonal structure. The $z$, $A$, and $B$ axes have  
     the same positions in all the three structures in
     Figs.~\ref{fig:structures} (a), (b), and (c).  The point group operations
     with the indices $x$, $y$, $z$, $a$, and $b$ are related to the depicted
     $x$, $y$, $z$, $A$, and $B$ axes, respectively. For instance, $C^-_{4z}$
     is a clockwise rotation of the lattice through $\frac{\pi}{2}$ radians
     about the $z$ axis, and $C_{2a}$ is a rotation through $\pi$ radians 
     about the $A$ axis. $\sigma_{da} = IC_{2a}$ ($I$ is the inversion) is a
     reflection in the plain containing the origin and being perpendicular
     to the $A$ axis.
 \label{fig:cartesian}
}
 \end{figure}

\begin{SCfigure*}[1][!]
\begin{minipage}[t]{.6\textwidth}
\centering
\includegraphics[width=.41\textwidth,angle=0]{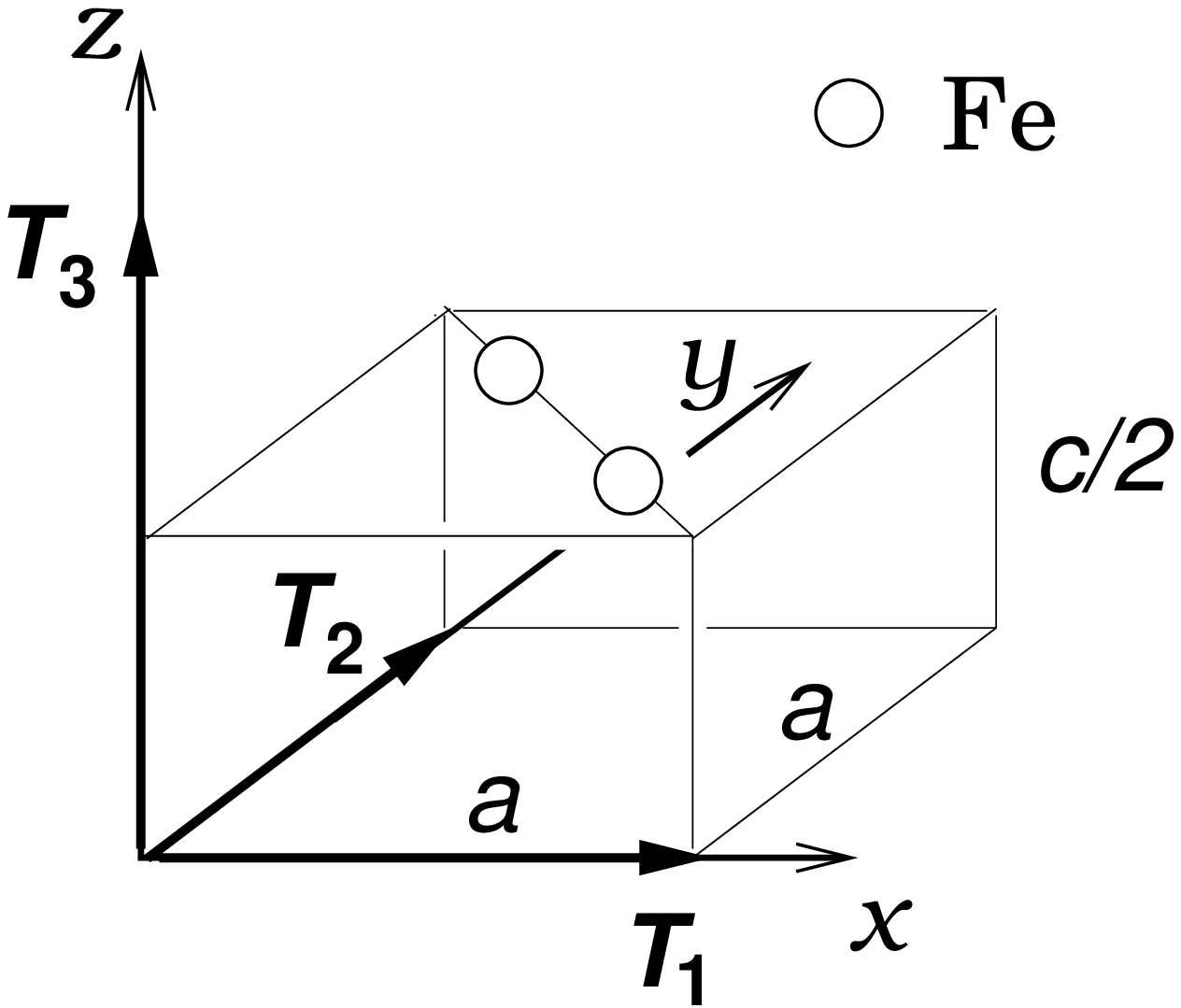}%
\begin{center}
(a)
\vspace{1cm}
\end{center}
\includegraphics[width=.73\textwidth,angle=0]{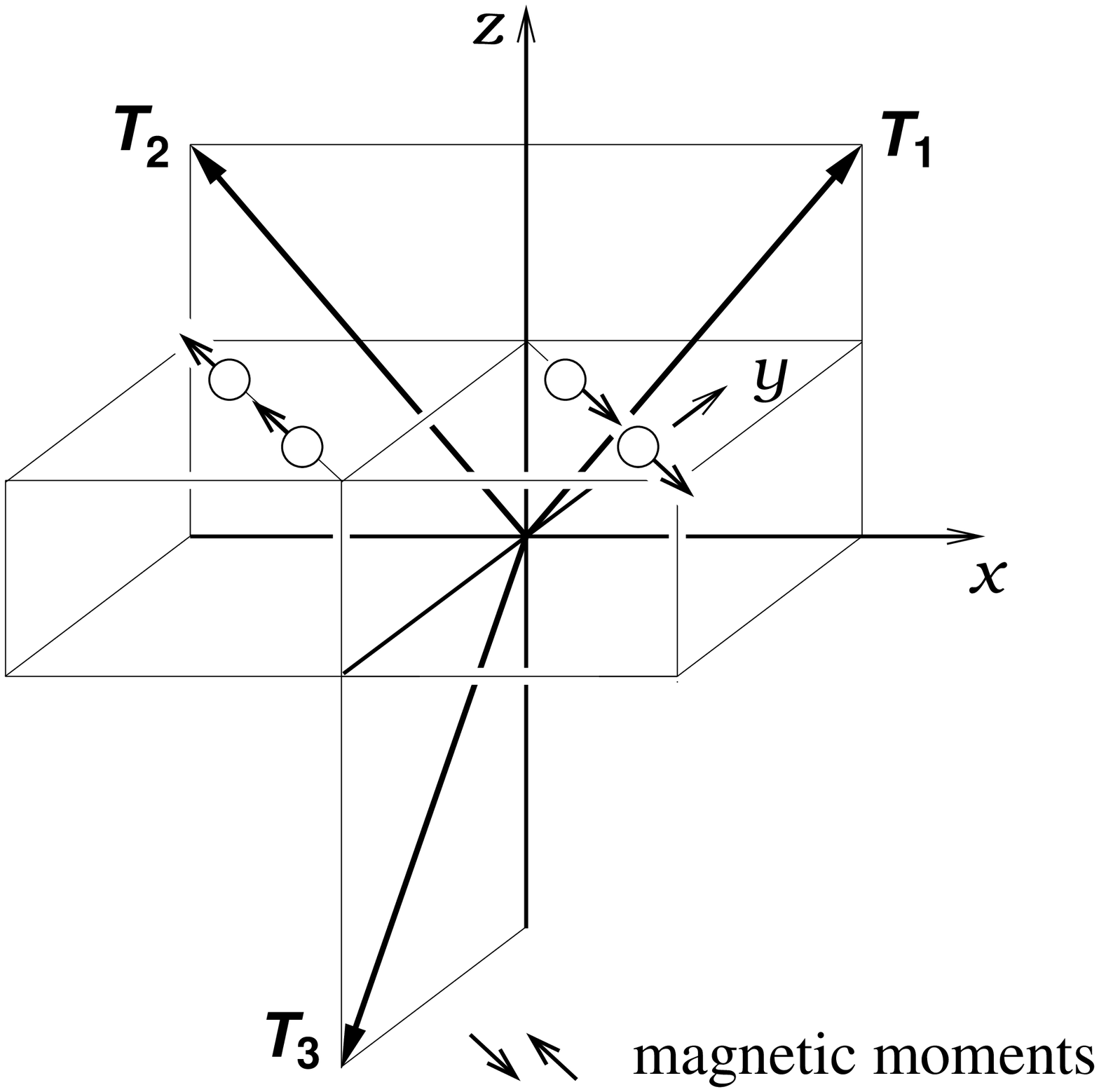}%
\begin{center}
(b)
\vspace{1cm}
\end{center}
\includegraphics[width=.9\textwidth,angle=0]{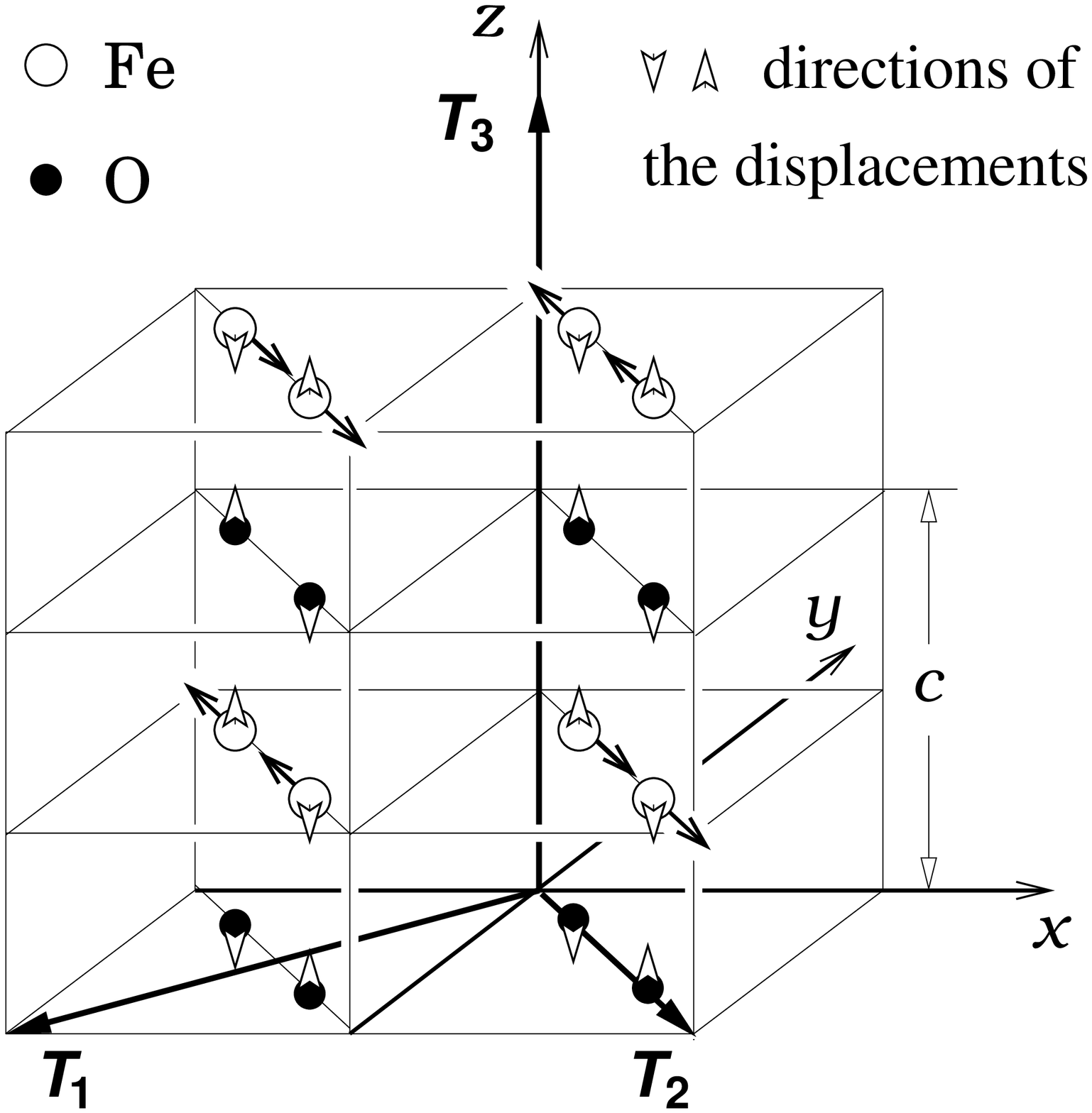}%
\begin{center}
(c)
\end{center}
\end{minipage}
\columnsep = -20cm
\caption{Coordinate systems and unit cells of three (magnetic) structures in
  LaFeAsO. For reasons of clarity, only the Fe atoms and
  in structure (c) also the O atoms are shown. 
  $a$ and $c$ denote the lengths of the sides in the tetragonal unit cell. 
  The coordinate systems define  
  the symmetry operations $\{R|pqr\}$ as used in this paper. They are written in
  the Seitz notation detailed in the textbook of Bradley and 
  Cracknell~\cite{bc}: $R$ stands for a point group operation and $pqr$ 
  denotes the subsequent translation. The point group operation $R$ is related 
  to the $x, y, z$ coordinate system as defined in
  Fig.~\ref{fig:cartesian}, and $pqr$ stands for the translation 
  $\bm t = p\bm T_1 + q\bm T_2 + r\bm T_3$ with the basic translations 
  $\bm T_i$ being different for the structures (a), (b), and (c).
  The origin of the coordinate systems is fixed for all the three structures 
  (a), (b), and (c).

  (a) The paramagnetic structure of LaFeAsO with the tetragonal space group 
  $P4/nmm$ (129).

  (b) The antiferromagnetic structure in the undistorted material with the
  orthorhombic space group $Imma$ (74).   

  (c) The antiferromagnetic structure in distorted LaFeAsO 
  with the ``allowed'' orthorhombic space group $Pnn2$ (34).
  This paper proposes the indicated displacements of the Fe and O atoms in $z$
  direction which are invariant under the magnetic group $M_{34} = Pnn2 +
  \{K|\frac{1}{2}\frac{1}{2}0\}Pnn2$.   
  Since $Pnn2$ is a subgroup of $Imma$,  
  the magnetic moments may have the same directions as in the undistorted
  structure (b). However, in the magnetic group $M_{34}$ the moments may also
  be canted within a plain perpendicular to the $z$ axis as depicted in
  Fig.~\ref{fig:ebene}.      
\label{fig:structures}
}
\end{SCfigure*}

The coordinates systems used in this paper are depicted in
Figs.~\ref{fig:cartesian} and~\ref{fig:structures}. First, we use the $x, y,
z$ coordinate systems as depicted in Fig.~\ref{fig:cartesian} which coincides
with the coordinates normally used in the literature on LaFeAsO, as, for
instance, in the tables of Ref.~\cite{clarina}. This coordinate system defines
the point group operations in the way described in Fig.~\ref{fig:cartesian}.


In the following we shall consider three structures in pure LaFeAsO whose unit
cells together with the basic translations $\bm T_1$, $\bm T_2$, and $\bm T_3$
of the corresponding Bravais lattices are given in
Fig.~\ref{fig:structures}. Fig.~\ref{fig:structures} (a) shows the
tetragonal primitive Bravais lattice $\Gamma_q$ of paramagnetic LaFeAsO above
$155 K$ which has the space group
$P4/nmm$~\cite{clarina,kamihara,prl_chen,nature_chen,wen,dong}.  In
Fig.~\ref{fig:structures} (b) the experimentally observed antiferromagnetic
structure in a hypothetically undistorted crystal is depicted, which has the
orthorhombic body-centered Bravais lattice $\Gamma^v_o$. Finally,
Fig.~\ref{fig:structures} (c) shows the antiferromagnetic structure in
distorted LaFeAsO below $137 K$ with the orthorhombic primitive Bravais
lattice $\Gamma_o$.


For the description of these three structures we need to resort to three
coordinate systems with unusual relationships.  While the basic translations
$\bm T_1$, $\bm T_2$, and $\bm T_3$ are adapted to the respective structure,
the origin of the $x, y, z$ coordinate systems and the directions of its axes
relative to the atoms are {\em fixed} for all the three structures in
Figs.~\ref{fig:structures}.  Therefore, we may compare directly the space
groups of the structures because the same point group operation in any
structure is notated by the same symbol.  As a consequence, the $A$ and $B$
axes (as given in Fig.~\ref{fig:cartesian}), but not the $x$ and $y$ axes, are
perpendicular to each other in the orthorhombic structures. Furthermore, this
choice of the $x, y, z$ axes requires a renaming of the point group operations
given in the tables in the textbook of Bradley and Cracknell~\cite{bc}, as
described in the notes to Tables~\ref{tab:rep129}, ~\ref{tab:rep74},
and~\ref{tab:rep34}.

\section{The undistorted magnetic structure}
\label{sec:undistorted}
In this section we show that the experimentally observed magnetic structure in
LaFeAsO is compatible with the symmetry of the Bloch functions of the magnetic
band denoted in Fig.~\ref{fig:bandstr} by the bold line. First, in the
following subsection we shall determine the magnetic group of the
experimentally observed magnetic structure under the assumption that there are
no structural distortions in the material in addition to the magnetostriction.

\subsection{The experimentally determined magnetic structure and its magnetic
  group}
\label{sec:74}
We show that the group $Imma = \Gamma^v_oD^{28}_{2h}$ (74) with the
orthorhombic body-centered Bravais lattice $\Gamma^v_o$ depicted in
Fig.~\ref{fig:structures} (b) is the space group of the experimentally
determined magnetic structure as it is given by Fig.~4 of
Ref.~\cite{clarina}. In Fig.~\ref{fig:structures} (b) the unit cell of
antiferromagnetic LaFeAsO is depicted where only the Fe atoms are
indicated since the magnetic moments are localized on the Fe
sites~\cite{clarina, nomura}. This Fig.~\ref{fig:structures} (b) is identical
with Fig.~4 of Ref.~\cite{clarina} because by application of the basic
translations $\bm T_1, \bm T_2, \bm T_3$ to the four Fe atoms in the unit cell
the magnetic structure in Fig.~4 of Ref.~\cite{clarina} can be constructed.

In a first step we show that the positions of the atoms of LaFeAsO are invariant
under the symmetry operations of $Imma$. They are clearly invariant under the
translations of $Imma$ since its basic translations $\bm T_1, \bm T_2, \bm
T_3$ as given in Fig.~\ref{fig:structures} (b) are also translations of the
paramagnetic structure depicted in Fig.~\ref{fig:structures} (a) [cf. the
following relation~\gl{eq:15}]. In order to show that the atoms of LaFeAsO are
also invariant under the rotations and reflections of $Imma$, it is sufficient
to consider the generating elements
\begin{equation} 
\label{eq:1}
\{C_{2z}|0\textstyle\frac{1}{2}\textstyle\frac{1}{2}\}, 
\{C_{2a}|\textstyle\frac{1}{2}0\textstyle\frac{1}{2}\},
\{I|\textstyle\frac{1}{2}\textstyle\frac{1}{2}0\}
\end{equation} 
of $Imma$ as they are given in Table 3.7 of Ref.~\cite{bc}. However, in this
paper they are written in the coordinate system defined by
Figs.~\ref{fig:cartesian} and~\ref{fig:structures} (b), cf. the notes to
Table~\ref{tab:rep74}. We may easily compare the space group $Imma$ with the
paramagnetic space group $P4/nmm$ by writing the translations of the
generating elements \gl{eq:1} in terms of the basic translations of the
Bravais lattice $\Gamma_q$ of $P4/nmm$ as depicted in
Fig.~\ref{fig:structures} (a). Using the relation
\begin{equation}
  \label{eq:15}
\left.
\begin{array}{lcl}
\bm T_1 & \rightarrow & \ \ \bm T_1 + \bm T_3\\
\bm T_2 & \rightarrow & -\bm T_1 + \bm T_3\\
\bm T_3 & \rightarrow & -\bm T_2 - \bm T_3.\\
\end{array}
\right\}
\ \Gamma^v_o\ \rightarrow\ \Gamma_q
\end{equation}
[as derived from Figs.~\ref{fig:structures} (a) and (b)] we get the symmetry
operations $\{C_{2z}|\overline{\frac{1}{2}}\overline{\frac{1}{2}}0\},
\{C_{2a}|\frac{1}{2}\overline{\frac{1}{2}}0\},$ and $\{I|001\}$, all of them
being in the tetragonal space group $P4/nmm$, see the symmetry operations
belonging to point $\Gamma$ in Table~\ref{tab:rep129}. Consequently, within
the coordinate systems defined in 
Figs.~\ref{fig:structures} (a) and (b), the group $Imma$ is a subgroup of
$P4/nmm$. The important implication is that the positions of the atoms of 
LaFeAsO are invariant under the symmetry operations of $Imma$.

Now we can show that also the magnetic structure in Fig.~\ref{fig:structures}
(b) is invariant under the generating elements \gl{eq:1} of $Imma$: One of the
four Fe atoms in the unit cell lies at the position $\bm p_1 = \frac{1}{2}\bm
T_1 + \frac{1}{4}\bm T_2 + \frac{1}{4}\bm T_3$. Applying on this atom, e.g.,
the rotation $C_{2z}$, we get an atom at the position $\bm p_2 =
\frac{1}{4}\bm T_2 - \frac{1}{4}\bm T_3$ with the false direction of the
magnetic moment. Then the associated translation $\bm t = \frac{1}{2}\bm T_2 +
\frac{1}{2}\bm T_3$, however, moves this atom to the position $\bm p_3 =
\frac{3}{4}\bm T_2 + \frac{1}{4}\bm T_3$ of an Fe atom with the correct
direction of the magnetic moment. In the same way, it can be shown that the
symmetry operation $\{C_{2z}|0\textstyle\frac{1}{2}\textstyle\frac{1}{2}\}$
leaves invariant the directions of the magnetic moments of the three other Fe
atoms in the unit cell. We get the same result for the other two generating
elements $\{C_{2a}|\textstyle\frac{1}{2}0\textstyle\frac{1}{2}\}$ and
$\{I|\textstyle\frac{1}{2}\textstyle\frac{1}{2}0\}$. Consequently, the group
$Imma$ is the space group of the experimentally determined magnetic structure
in LaFeAsO.

In addition, the magnetic structure depicted in Fig.~\ref{fig:structures} (b)
is invariant under the anti-unitary operator $\{K|\frac{1}{2}\frac{1}{2}0\}$,
where $K$ denotes the operator of time inversion.  First, $K$ reverses all the
magnetic moments (and leaves invariant the positions of the atoms). Then the
associated translation $\bm t = \frac{1}{2}\bm T_1 + \frac{1}{2}\bm T_2$ moves
the Fe atoms to positions with the original directions of the magnetic
moments. Also $\bm t$ leaves invariant the positions of the atoms because it
is a lattice translation (namely $\bm T_3$) in the paramagnetic lattice
depicted in Fig.~\ref{fig:structures} (a).  Hence, the magnetic group $M$ of
the experimentally determined magnetic structure in undistorted LaFeAsO as
depicted in Fig.~\ref{fig:structures} (b) may be written as
\begin{equation}
  \label{eq:3}
  M = Imma + \{K|\textstyle\frac{1}{2}\frac{1}{2}0\}Imma.
\end{equation}

\subsection{The symmetry of the Bloch functions of the magnetic band}
\label{sec:bf}
The energy band of LaFeAsO denoted in Fig.~\ref{fig:bandstr} by the bold line 
is characterized by the representations
\begin{equation}
  \label{eq:2}
\Gamma^-_2, \Gamma^+_5; X_1, X_1; M_3, M_4; A_3, A_2; Z^+_1, Z^-_5; R_1, R_1.
\end{equation}
Folding this energy band into the Brillouin zone of the space group $Imma$ of
the antiferromagnetic structure in the undistorted crystal depicted in
Fig.~\ref{fig:structures} (b), the representations~\gl{eq:2} of the Bloch
functions transform as
\begin{equation}
  \label{eq:7}
\begin{array}{lclp{.3cm}lcl}
\Gamma^-_2 & \rightarrow & \underline{\Gamma^-_3} && 
\Gamma^+_5 & \rightarrow & \underline{\Gamma^+_2} + \Gamma^+_4\\[.2cm] 
M_3 & \rightarrow & X^+_3 + \underline{X^-_1}&&  
M_4 & \rightarrow & \underline{X^+_4} + X^-_4\\[.2cm]
A_3 & \rightarrow & \underline{\Gamma^+_3} + \Gamma^-_1&&  
A_2 & \rightarrow & \Gamma^+_2 + \underline{\Gamma^-_2}\\[.2cm]
Z^+_1 & \rightarrow & \underline{X^+_1}&&  
Z^-_5 & \rightarrow & X^-_2 + \underline{X^-_4}\\[.2cm]  
R_1, X_1 & \rightarrow & \underline{T_1} + T_1&&  
R_1, X_1 & \rightarrow & \underline{T_1} + T_1\\[.2cm]  
\end{array}
\end{equation}
see Table~\ref{tab:falten}. The underlined representations form a band listed
in Table~\ref{tab:wf74}, namely band 2 in Table~\ref{tab:wf74} (b).  Hence,
the Bloch functions of this band can be unitarily transformed into Wannier
functions that are
\begin{itemize}
\item as well localized as possible, 
\item centered at the Fe atoms,
\item and symmetry-adapted to the magnetic group $M$ in Eq.~\gl{eq:3}, 
\end{itemize}
see the notes to Table~\ref{tab:wf74}. For this reason, this band is called
``magnetic band'' related to the magnetic group $M$. The NHM predicts that the
electrons of this partly filled band may lower their Coulomb correlation
energy by activating an exchange mechanism producing a magnetic structure with
the magnetic group $M$ and with the magnetic moments lying on the Fe sites,
i.e., by producing the experimentally determined magnetic
structure~\cite{enhm,ea,ef}. In this sense we say that the symmetry of the
Bloch functions is ``compatible'' with the experimentally determined magnetic
structure.

Table~\ref{tab:wf74} lists all the possible magnetic bands related to the
magnetic group $M$ in Eq.~\gl{eq:3}. By this table one can make sure that the
symmetry of the Bloch functions in relation~\gl{eq:7} corresponds only to band
2 in Table~\ref{tab:wf74} (b). That is, the magnetic moments in LaFeAsO may
only be situated at the Fe sites.

\section{Stability of a magnetic structure}
\label{sec:stability}
The magnetic band denoted in Fig.~\ref{fig:bandstr} by the bold line is
related to the magnetic group $M$ in Eq.~\gl{eq:3}. However, this structure
cannot exist in the undistorted material because the space group $Imma$ does
not possess suitable representations.  This shall be substantiated in the
following.

Let be 
\begin{equation}
  \label{eq:14}
  M = S + \{K|\bm t\}S
\end{equation}
the magnetic group of a given antiferromagnetic structure, with $S$ denoting
the space group of this structure and $\{K|\bm t\}$ being an anti-unitary
operator leaving invariant the magnetic structure. $K$ still denotes the
operator of time inversion.

The operator $K$ reverses the magnetic moments in the antiferromagnetic ground
state $|G\rangle$, so
\begin{equation}
  \label{eq:6}
\overline{|G}\rangle = K|G\rangle
\end{equation}
is the state with the opposite directions of the magnetic moments which
clearly is different from $|G\rangle$. Within the NHM, both states
$\overline{|G}\rangle$ and $|G\rangle$ are eigenstates of a Hamiltonian
denoted in Ref.~\cite{ea} by $\widetilde H$ which commutes with $K$.
Hence, $\overline{|G}\rangle$ and $|G\rangle$ belong to an {\em irreducible}
two-dimensional corepresentation $\widetilde D^{M}$ of the group
\begin{equation}
  \label{eq:8}
\widetilde M = M + KM  
\end{equation}
of the Hamiltonian $\widetilde H$, see Ref.~\cite{ea}, Sec.~III.C.

However, when we restrict ourselves to the symmetry operations $P$ of the
subgroup $M$ of $\widetilde M$, then $\widetilde D^{M}$ must subduce a {\em
  one-dimensional} corepresentation $D^M$ of $M$, that is,
\begin{equation}
  \label{eq:19}
P|G\rangle = c\cdot|G\rangle \text{ for } P\in M,
\end{equation}
where $|c|$ = 1.  In particular, this Eq.~\gl{eq:19} is satisfied for the
anti-unitary operator $\{K|\bm t\}$,
\begin{equation}
  \label{eq:20}
\{K|\bm t\}|G\rangle = c\cdot|G\rangle,
\end{equation}
where still $|c|$ = 1.

A {\em stable} magnetic state $|G\rangle$ may exist if $\widetilde M$
possesses at least one suitable corepresentation $\widetilde D^{M}$. From this
condition it follows the
\begin{thm}
\label{theorem}
A stable magnetic state with the
space group $S$ can exist if $S$ has at least one one-dimensional
single-valued representation $D$
\begin{enumerate} 
\item 
following case (a) with respect to the magnetic group 
$S + \{K|\bm t\}S$ and
\item 
following case (c) with
respect to the magnetic group $S + KS$.
\end{enumerate}
\end{thm}
The cases (a) and (c) are defined in Eqs.\ (7.3.45) and (7.3.47),
respectively, of Ref.~\cite{bc}. They determine the dimension of the
corepresentations of the magnetic groups $S + \{K|\bm t\}S$ and $S + KS$,
respectively, which are derived~\cite{bc} from the representation $D$ of
$S$. The one-dimensional representation $D$ may satisfy the second condition
(ii) only if $D$ has non-real characters.

This Theorem~\ref{theorem} was proposed in Ref.~\cite{ea} and written
down in the present form in Ref.~\cite{josla2cuo4}. It can be understood
following the theory of corepresentations as given in Sec.~7.3. of the
textbook of Bradley and Cracknell~\cite{bc}: Let $D$ be a representation of $S$
satisfying the first condition (i) of the theorem, then we may derive from $D$
the one-dimensional corepresentation $D^M$ of $M$ given by Eq.~(7.3.45) of
Ref.~\cite{bc} (with $A = \{K|\bm t\}$ and {\bf N} $ = 1$).

Then we may derive from $D^M$ the two-dimensional corepresentation $\widetilde
D^{M}$ of $\widetilde M$ as it is given in Eq.~(7.3.17) of Ref.~\cite{bc},
where now $A = K$. [Bradley and Cracknell assume that the subgroup $G$ in
their Eq.~(7.3.11) does not contain anti-unitary elements; for the derivation
of their Eq.~(7.3.17), however, these authors do not make use of this
assumption.] If $D$ satisfies additionally the second condition (ii) of the
theorem, then the corepresentation $\widetilde D^{M}$ has just the required
properties, cf. the Auxiliary Publication, citation 9, of Ref.~\cite{ea}.

In the present paper the cases (a) and (c) are determined (in
Tables~\ref{tab:rep74} and~\ref{tab:rep34}) by the relatively straightforward
equation (7.3.51) of Ref.~\cite{bc},
\begin{equation}
  \label{eq:18}
\sum_{B\in C} \chi (B^2) = \left\{ 
\begin{array}{cl}
+|S| & \text{in case (a)}\\
-|S| & \text{in case (b)}\\
0 & \text{in case (c)}.\\
\end{array}
 \right.
\end{equation}
The sum runs over the symmetry operations $B$ in the left cosets $C = \{K|\bm
t\}S$ and $C = KS$, respectively, of the magnetic groups. $\chi (B^2)$ is the
character of $B^2$ in the representation $D$ and $|S|$ denotes, as usual, the
order of $S$.

Table~\ref{tab:rep74} shows that the space group $Imma$ does not possess
one-dimensional representations with non-real characters. Consequently, all
the one-dimensional representations of $Imma$ follow case (a) with respect to
the magnetic group $Imma + KImma$ and hence, stable magnetic structures with
the space group $Imma$ do not exist.

\section{Allowed space groups in antiferromagnetic $\text{LaFeAsO}$ }
\label{sec:subgroups}
As shown in the preceding Sec.~\ref{sec:stability}, the magnetic structure
depicted in Fig.~\ref{fig:structures} (b) is not stable because its space
group $Imma$ does not possess suitable representations. However, this
structure may be stabilized by a (small) distortion of the crystal that turns
the space group $Imma$ into a new space group possessing representations that
allow the formation of a stable magnetic structure. This new space group shall
be named an ``allowed'' space group in LaFeAsO. In this context we assume that
the distortions of the crystal are so small that the magnetic structure in the
distorted material differs hardly from the structure depicted in
Fig.~\ref{fig:structures} (b) because the symmetry of the Bloch functions of
the magnetic band is compatible with the magnetic group $M$ in
Eq.~\ref{eq:3}. For that reason we assume that the magnetic structure depicted
in Fig.~\ref{fig:structures} (b) is invariant under an allowed space group,
i.e., we assume that an allowed space group is a subgroup of $Imma$.

In this section we look for all the allowed space groups in antiferromagnetic
LaFeAsO by considering all the orthorhombic primitive and monoclinic primitive
space groups in Table 3.7 of Ref.~\cite{bc} with at least four point group
elements. This is a laborious task because all the conceivable translations and
rotations of the coordinate systems given in Table 3.7 of Ref.~\cite{bc}
should be regarded.

The coordinate system of the magnetic structure in LaFeAsO in an orthorhombic
primitive or monoclinic primitive Bravais lattice is depicted in
Fig.~\ref{fig:structures} (c). The positions of the atoms and the magnetic
structure are invariant under two anti-unitary operations of the form $\{K|\bm
t\}$, namely under $\{K|\frac{1}{2}\frac{1}{2}0\}$ and
$\{K|00\frac{1}{2}\}$. Hence, we must look for space groups $S$ possessing
one-dimensional representations following case (c) with respect to the
magnetic group $S + KS$ and
case (a) with respect either to $S + \{K|\frac{1}{2}\frac{1}{2}0\}S$ or to
$S + \{K|00\frac{1}{2}\}S$, see Theorem~\ref{theorem}.

We found the two allowed space groups $Pnn2 = \Gamma_oC^{10}_{2v}$ (34) and
$P_1P_1P_1 = \Gamma_oD^4_2$ (19) and shall present them in the following two
Secs.~\ref{sec:34} and~\ref{sec:19}. In Sec.~\ref{sec:distortion} we will show
that $Pnn2$ clearly is the space group of LaFeAsO.

\subsection{The space group $Pnn2$}
\label{sec:34}
The space group $Pnn2$ has the orthorhombic-primitive Bravais lattice
$\Gamma_o$ and can be defined by the two generating elements
\begin{equation}
  \label{eq:9}
\{\sigma_{da}|0\textstyle\frac{1}{2}\textstyle\frac{1}{2}\}\ \text{and}\ 
\{\sigma_{db}|\textstyle\frac{1}{2}\textstyle\frac{1}{2}\textstyle\frac{1}{2}\},
\end{equation}
cf. Table 3.7 of Ref.~\cite{bc} and the notes to Table~\ref{tab:rep34}. The
symmetry operations~\gl{eq:9} are given in the coordinate system defined in
Figs.~\ref{fig:cartesian} and~\ref{fig:structures} (c). In order to show that
$Pnn2$ is a subgroup of the group $Imma$ (with the Bravais lattice
$\Gamma^v_o$) we write the translations of the generating elements \gl{eq:9}
in terms of the basic translations of the group $Imma$ as given in
Fig.~\ref{fig:structures} (b). Using the relation
\begin{equation}
  \label{eq:13}
\left.
\begin{array}{lcl}
\bm T_1 & \rightarrow & \bm T_2 + \bm T_3\\
\bm T_2 & \rightarrow & \bm T_1 + \bm T_3\\
\bm T_3 & \rightarrow & \bm T_1 + \bm T_2,\\
\end{array}
\right\}
\ \Gamma_o\ \rightarrow\ \Gamma^v_o
\end{equation}
[as deduced from Figs.~\ref{fig:structures} (b) and (c)], we get the operations
$\{\sigma_{da}|1\textstyle\frac{1}{2}\textstyle\frac{1}{2}\}$ and
$\{\sigma_{db}|111\}$, both belonging to the group $Imma$, see the symmetry
operations belonging to point $\Gamma$ in Table~\ref{tab:rep74}. Consequently,
within the coordinate systems of Fig.~\ref{fig:structures}, $Pnn2$ is a
subgroup of $Imma$.

In contrast to the group $Imma$, the group $Pnn2$ possesses at point $S$
one-dimensional representations satisfying the conditions (i) and (ii) of
Theorem~\ref{theorem} for the magnetic groups 
\begin{equation}
  \label{eq:10}
  M_{34} = Pnn2 + 
      \{K|\textstyle\frac{1}{2}\textstyle\frac{1}{2}0\}Pnn2    
\end{equation}
and $Pnn2 + KPnn2$, respectively, see Table~\ref{tab:rep34}. Further, the
little group of $S$ comprises the whole space group. Hence, $Pnn2$ may be the
space group of a {\em stable} magnetic structure. A stable antiferromagnetic
ground state $|G\rangle$ is basis function of a one-dimensional
corepresentation of $M_{34}$ and $|G\rangle$ and the time inverted state
$\overline{|G}\rangle = K|G\rangle$ form basis functions of an {\em
  irreducible} two-dimensional corepresentation of the group
$$
\widetilde M_{34} = M_{34} + KM_{34}
$$
which might be determined by Eq.~(7.3.17) of Ref.~\cite{bc}. [As an
example, the analogous corepresentation for the spin-density-wave state in
chromium is explicitly given in Ref.~\cite{ea}.]

\subsection{The space group $P_1P_1P_1$}
\label{sec:19}
A second allowed space group is the group $P_1P_1P_1$
with the generating elements
\begin{equation}
  \label{eq:11}
\{C_{2a}|\textstyle\frac{1}{2}0\textstyle\frac{1}{2}\} \text{ and }
\{C_{2b}|\textstyle\frac{1}{2}\textstyle\frac{1}{2}0\},
\end{equation}
written again within the coordinate systems of Figs.~\ref{fig:cartesian}
and~\ref{fig:structures} (c). We do not describe this group but only report
the result that a magnetic structure with the magnetic group
\begin{equation}
  \label{eq:12}
  M_{19} = P_1P_1P_1 + 
      \{K|\textstyle\frac{1}{2}\textstyle\frac{1}{2}0\}P_1P_1P_1    
\end{equation}
could be stable in LaFeAsO. However, antiferromagnetic LaFeAsO does
not possess this space group, see the following
Sec.~\ref{sec:distortion}.
\section{Distortion of the crystal}
\label{sec:distortion}
The antiferromagnetic structure in undistorted LaFeAsO has the space group
$Imma$. Hence, in the undistorted antiferromagnetic material the nonadiabatic
Hamiltonian $H^n$ [as defined in Eq.~(2.15) of Ref.~\cite{enhm}] would
commute with all the symmetry operations of $Imma$. However, as shown in the
preceding sections, $Imma$ is not an allowed space group but has two allowed
subgroups, namely $Pnn2$ and $P_1P_1P_1$. Thus, either $Pnn2$ or $P_1P_1P_1$,
but not $Imma$ may be the space group of the antiferromagnetic structure in
LaFeAsO. Consequently, the nonadiabatic Hamiltonian $H^n$ commutes with the
symmetry operations $P$ of one of the allowed space groups $S$, but does not
commute with those symmetry operations that belong to $Imma$, but do not
belong to $S$,
\begin{equation}
  \label{eq:16}
\begin{array}{lccll}
\ [H^n,P] & = & 0 & \text{ for } & P \in S\\
\ [H^n,P] & \neq & 0 & \text{ for } & P \in (Imma - S).\\
\end{array}
\end{equation}
For that reason, the Fe and O atoms in LaFeAsO are slightly shifted from their
positions in the space group $Imma$.

In order to show this, we write the eight symmetry operations of $Imma$ as
given in Table~\ref{tab:rep74} for point $\Gamma$ in terms of the basic
translations $\bm T_1$, $\bm T_2$, and $\bm T_3$ of the orthorhombic primitive
lattice $\Gamma_o$ in Fig.~\ref{fig:structures} (c),
\begin{equation}
  \label{eq:4}
\begin{array}{ll}
\{E|000\}, & \{E|\frac{1}{2}\frac{1}{2}\frac{1}{2}\},\\
\{C_{2z}|\frac{1}{2}00\}, & \{C_{2z}|0\frac{1}{2}\frac{1}{2}\},\\
\{C_{2a}|0\frac{1}{2}0\}, & \{C_{2a}|\frac{1}{2}0\frac{1}{2}\},\\
\{C_{2b}|\frac{1}{2}\frac{1}{2}0\}, & \{C_{2b}|00\frac{1}{2}\},\\
\{I|\frac{1}{2}\frac{1}{2}0\}, & \{I|00\frac{1}{2}\},\\
\{\sigma_{z}|\frac{1}{2}0\frac{1}{2}\}, & \{\sigma_{z}|0\frac{1}{2}0\},\\
\{\sigma_{da}|\frac{1}{2}00\}, &
\{\sigma_{da}|0\frac{1}{2}\frac{1}{2}\},\\ 
\{\sigma_{db}|000\}, & \text{ and }
\{\sigma_{db}|\frac{1}{2}\frac{1}{2}\frac{1}{2}\},\\ 
\end{array}
\end{equation}
cf. relation~\gl{eq:13}. For each symmetry operation from $\Gamma^v_o$ we get
two operations in $\Gamma_o$ because the operation
$\{E|\frac{1}{2}\frac{1}{2}\frac{1}{2}\}$ in $\Gamma_o$ is a lattice
translation in $\Gamma^v_o$ ($E$ is the identity). 

The antiferromagnetic structure with the space group $Imma$ is invariant under
the sixteen symmetry operations~\gl{eq:4} now written within the Bravais
lattice $\Gamma_o$ of the structure in Fig.~\ref{fig:structures} (c). By
inspection of Fig.~\ref{fig:structures} (c) we find under these sixteen
operations four operations, namely
\begin{equation}
  \label{eq:17}
\begin{array}{ll}
\{E|000\}, & \{C_{2z}|\frac{1}{2}00\},\\
\{\sigma_{da}|0\frac{1}{2}\frac{1}{2}\}, & \text{ and } 
\{\sigma_{db}|\frac{1}{2}\frac{1}{2}\frac{1}{2}\},\\ 
\end{array}
\end{equation}
leaving invariant the {\em distorted}\ crystal as depicted in
Fig.~\ref{fig:structures} (c).  These four operations form (together with the
translations) a group, namely the allowed space group $Pnn2$. Hence, the
experimentally observed~\cite{clarina} displacements of the Fe and O atoms
``realize'' the space group $Pnn2$ in the sense that they produce the correct
commutation properties [given by Eq.~\gl{eq:16}] of the nonadiabatic
Hamiltonian $H^n$ and stabilize in this way the magnetic structure in LaFeAsO.
Together with the anti-unitary operator $\{K|\frac{1}{2}\frac{1}{2}0\}$ (which
also leaves invariant the distorted structure), the symmetry
operations~\gl{eq:17} form the magnetic group $M_{34}$ in Eq.~\gl{eq:10}.

\begin{figure}[t]
   \includegraphics[width=.3\textwidth,angle=0]{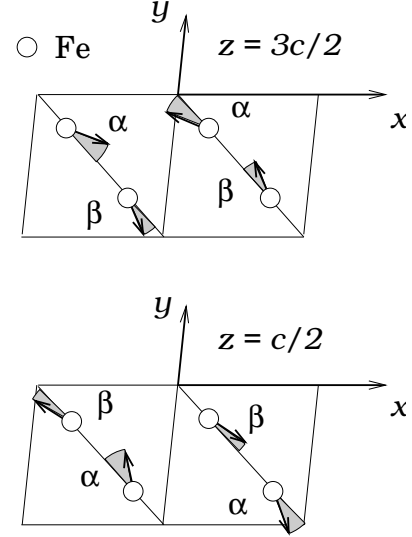}
   \caption{The two layers of the Fe atoms in the unit cell of the distorted
     antiferromagnetic structure depicted in Fig.~\ref{fig:structures} (c).
     The $z$ axis is normal to the paper. For clarity, the small displacements of
     the Fe atoms (in $z$ direction) are not indicated. The 
     $x, y, z$ coordinate system coincides with the $x, y, z$ coordinates in
     Fig.~\ref{fig:structures} (c). Within distorted LaFeAsO, 
     the magnetic moments as depicted in Fig.~\ref{fig:structures} (c) may be
     rotated by arbitrary angles $\alpha$ and $\beta$ within a plain
     perpendicular to the $z$ axis because also this canted structure is
     invariant under the magnetic group $M_{34} = Pnn2 +
     \{K|\frac{1}{2}\frac{1}{2}0\}Pnn2$.  $\alpha$ and $\beta$ may be different 
     and both angles are likely small because the symmetry of the Bloch 
     functions of the magnetic band is compatible with the magnetic group $M$ of
     the magnetic structure depicted in Fig.~\ref{fig:structures} (b). 
 \label{fig:ebene}
}
 \end{figure}

In the magnetic group $M_{34}$ the magnetic moments may be canted by two
arbitrary angles $\alpha$ and $\beta$ as depicted in Fig.~\ref{fig:ebene}
(while such canted moments are not invariant under $Imma$). Both angles are
likely small because the symmetry of the Bloch functions of the magnetic band
is compatible with the magnetic group $M$ [in Eq.~\gl{eq:3}] of the structure
depicted in Fig.~\ref{fig:structures} (b).

For the second allowed space group $P_1P_1P_1$, on the other hand, we have not
found any simple distortion of the lattice realizing this group.


\section{Conclusions}

Within the nonadiabatic Heisenberg model (NHM), the situation in the magnetic
band of LaFeAsO (denoted in Fig.~\ref{fig:bandstr} by the bold line) is
characterized by two different group-theoretical phenomena. On the one hand,
the magnetic band (i.e., the symmetry of the Bloch functions of this band) is
related to the magnetic group $M$ of the experimentally
observed~\cite{clarina, nomura,kitao,nakai} magnetic structure in LaFeAsO and,
on the other hand, any structure with the space group $M$ is unstable. These
two phenomena are discussed in separated subsections~\ref{sec:energy} and
\ref{sec:stablems}.

\subsection{Nonadiabatic condensation energy} 
\label{sec:energy}
The nonadiabatic Heisenberg model (NHM) is defined by three postulates on the
Coulomb correlation energy in narrow, partly filled bands~\cite{enhm}. A direct
consequence of these postulates is the existence of the ``nonadiabatic
condensation energy'' $\Delta E$ defined by Eq.~(2.20) of
Ref.~\cite{enhm}. The electrons at the Fermi level may lower their
Coulomb correlation energy by $\Delta E$ by occupying an atomic-like state as
defined by Mott~\cite{mott} and Hubbard~\cite{hubbard}: the electrons occupy
localized states as long as possible and perform their band motion by hopping
from one atom to another.  In the present approach, the atomic-like state is
consistently described in terms of symmetry-adapted and optimally localized
Wannier functions and is precisely defined by an equation in the nonadiabatic
system, namely by Eq.~(2.19) of Ref.~\cite{enhm}. In case the considered
partly filled band is a magnetic band related to a magnetic group $M$ (as
considered in this paper), the electrons of this band may gain the energy
$\Delta E$ by activating an exchange mechanism producing a magnetic structure
with the magnetic group $M$~\cite{enhm,ea,ef}.

In former papers we could show that evidently the magnetic states in
Cr~\cite{ea}, Fe~\cite{ef}, La$_2$CuO$_4$~\cite{josla2cuo4}, and
YBa$_2$Cu$_3$O$_6$~\cite{ybacuo6} are connected with narrow, partly filled
magnetic bands in the band structures of the respective materials. That is,
there is evidence that the nonadiabatic condensation energy $\Delta E$ is
responsible for the occurrence of the magnetic states in these materials.

Also LaFeAsO possesses a magnetic band related to the magnetic group $M = Imma
+ \{K|\frac{1}{2}\frac{1}{2}0\}$ of the experimentally observed magnetic
structure. This finding suggests that, first, the electrons of the magnetic
band of LaFeAsO perform the atomic-like motion as defined within the NHM by
Eq.~(2.19) of Ref.~\cite{enhm} and that, secondly, the nonadiabatic
condensation energy $\Delta E$ is responsible for the magnetic state.

\subsection{Stable  magnetic structures} 
\label{sec:stablems}
According to Theorem~\ref{theorem}, not all the magnetic structures may be
stable within the NHM. This theorem led already to an understanding of the
really existing magnetic groups in Cr~\cite{ea},
La$_2$CuO$_4$~\cite{josla2cuo4}, and YBa$_2$Cu$_3$O$_6$~\cite{ybacuo6}. In the
present paper, this theorem is the key for an understanding of the distortion
(in addition to the magnetostriction) of antiferromagnetic LaFeAsO.

Theorem~\ref{theorem} defines ``allowed'' space groups belonging to {\em
  stable} magnetic structures. Though the magnetic band of LaFeAsO is related
to the magnetic group $M = Imma + \{K|\frac{1}{2}\frac{1}{2}0\}$, the space
group $Imma$ is not allowed. On the other hand, allowed space groups in
LaFeAsO are the two groups $Pnn2$ and $P_1P_1P_1$. Consequently, the magnetic
structure may be stabilized by a (slight) change of the atomic sites in such a
way that the space group $Imma$ is turned into one of the allowed space
groups. Briefly speaking, the structural distortion ``realizes'' the space
group allowed in LaFeAsO.

The displacements of the Fe and O atoms as proposed in this paper are depicted
in Fig.~\ref{fig:structures} (c). They have the space group $Pnn2$ and realize
exactly this allowed space group. These displacements coincide in essence with
the theoretically determined~\cite{yildirim} and experimentally
observed~\cite{clarina} structural distortions of LaFeAsO.
 
However, we could not confirm the space groups $Cmma$ and $P112n$ as they have
been reported for the space group of antiferromagnetic LaFeAsO in the
undistorted~\cite{nomura} and the distorted~\cite{clarina} crystal,
respectively. Nevertheless, using the space group $P112n$, Clarina de la Cruz
et al.~\cite{clarina} discovered the important displacements of the Fe and O
atoms in $z$ direction. The directions of the displacements depicted in
Fig.~\ref{fig:structures} (c) do not essentially differ from those given in
Table 2 of Ref.~\cite{clarina}: while they coincide absolutely within
the Fe or O layers, Table 2 of Ref.~\cite{clarina} suggests that the
displacements are periodic with $\bm T_3/2$. As depicted in
Fig.~\ref{fig:structures} (c), the displacements proposed in this paper
alternate their directions at an interval of $\bm T_3/2$. However, in the
space group $Pnn2$ the amounts of the displacements in up and down direction
may be principally different because they are not connected by symmetry
operations.

\begin{acknowledgements}
  We wish to thank Volker Blum from the aims team of the Fritz-Haber-Institut
  der Max-Planck-Gesellschaft in Berlin for extending the aims program by an
  output of the eigenvectors which enabled us to determine the symmetry of the
  Bloch functions in the band structure of LaFeAsO. We are indebted to
  Franz-Werner Gergen from the EDV group of the Max-Planck-Institut f\"ur
  Metallforschung in Stuttgart for his assistance in getting to run the
  computer programs needed for this work, and we thank Ernst Helmut Brandt for
  valuable discussion and Ove Jepsen for initiating this paper.
\end{acknowledgements}

\FloatBarrier


\begin{thebibliography}{23}%
\makeatletter
\providecommand \@ifxundefined [1]{%
 \@ifx{#1\undefined}
}%
\providecommand \@ifnum [1]{%
 \ifnum #1\expandafter \@firstoftwo
 \else \expandafter \@secondoftwo
 \fi
}%
\providecommand \@ifx [1]{%
 \ifx #1\expandafter \@firstoftwo
 \else \expandafter \@secondoftwo
 \fi
}%
\providecommand \natexlab [1]{#1}%
\providecommand \enquote  [1]{``#1''}%
\providecommand \bibnamefont  [1]{#1}%
\providecommand \bibfnamefont [1]{#1}%
\providecommand \citenamefont [1]{#1}%
\providecommand \href@noop [0]{\@secondoftwo}%
\providecommand \href [0]{\begingroup \@sanitize@url \@href}%
\providecommand \@href[1]{\@@startlink{#1}\@@href}%
\providecommand \@@href[1]{\endgroup#1\@@endlink}%
\providecommand \@sanitize@url [0]{\catcode `\\12\catcode `\$12\catcode
  `\&12\catcode `\#12\catcode `\^12\catcode `\_12\catcode `\%12\relax}%
\providecommand \@@startlink[1]{}%
\providecommand \@@endlink[0]{}%
\providecommand \url  [0]{\begingroup\@sanitize@url \@url }%
\providecommand \@url [1]{\endgroup\@href {#1}{\urlprefix }}%
\providecommand \urlprefix  [0]{URL }%
\providecommand \Eprint [0]{\href }%
\providecommand \doibase [0]{http://dx.doi.org/}%
\providecommand \selectlanguage [0]{\@gobble}%
\providecommand \bibinfo  [0]{\@secondoftwo}%
\providecommand \bibfield  [0]{\@secondoftwo}%
\providecommand \translation [1]{[#1]}%
\providecommand \BibitemOpen [0]{}%
\providecommand \bibitemStop [0]{}%
\providecommand \bibitemNoStop [0]{.\EOS\space}%
\providecommand \EOS [0]{\spacefactor3000\relax}%
\providecommand \BibitemShut  [1]{\csname bibitem#1\endcsname}%
\let\auto@bib@innerbib\@empty
\bibitem [{\citenamefont {Nomura}\ \emph {et~al.}(2008)\citenamefont {Nomura},
  \citenamefont {Kim}, \citenamefont {Kamihara}, \citenamefont {Hirano},
  \citenamefont {Sushko}, \citenamefont {Kato}, \citenamefont {Takata},
  \citenamefont {Shluger},\ and\ \citenamefont {Hosono}}]{nomura}%
  \BibitemOpen
  \bibfield  {author} {\bibinfo {author} {\bibfnamefont {T.}~\bibnamefont
  {Nomura}}, \bibinfo {author} {\bibfnamefont {S.~W.}\ \bibnamefont {Kim}},
  \bibinfo {author} {\bibfnamefont {Y.}~\bibnamefont {Kamihara}}, \bibinfo
  {author} {\bibfnamefont {M.}~\bibnamefont {Hirano}}, \bibinfo {author}
  {\bibfnamefont {P.~V.}\ \bibnamefont {Sushko}}, \bibinfo {author}
  {\bibfnamefont {K.}~\bibnamefont {Kato}}, \bibinfo {author} {\bibfnamefont
  {M.}~\bibnamefont {Takata}}, \bibinfo {author} {\bibfnamefont {A.~L.}\
  \bibnamefont {Shluger}}, \ and\ \bibinfo {author} {\bibfnamefont
  {H.}~\bibnamefont {Hosono}},\ }\href@noop {} {\bibfield  {journal} {\bibinfo
  {journal} {Supercond. Sci. Technol.}\ }\textbf {\bibinfo {volume} {21}},\
  \bibinfo {pages} {125028} (\bibinfo {year} {2008})}\BibitemShut {NoStop}%
\bibitem [{\citenamefont {Kitao}\ \emph {et~al.}(2008)\citenamefont {Kitao},
  \citenamefont {Kobayashi}, \citenamefont {Higashitaniguchi}, \citenamefont
  {Saito}, \citenamefont {Kamihara}, \citenamefont {Hirano}, \citenamefont
  {Mitsui}, \citenamefont {Hosono},\ and\ \citenamefont {Seto}}]{kitao}%
  \BibitemOpen
  \bibfield  {author} {\bibinfo {author} {\bibfnamefont {S.}~\bibnamefont
  {Kitao}}, \bibinfo {author} {\bibfnamefont {Y.}~\bibnamefont {Kobayashi}},
  \bibinfo {author} {\bibfnamefont {S.}~\bibnamefont {Higashitaniguchi}},
  \bibinfo {author} {\bibfnamefont {M.}~\bibnamefont {Saito}}, \bibinfo
  {author} {\bibfnamefont {Y.}~\bibnamefont {Kamihara}}, \bibinfo {author}
  {\bibfnamefont {M.}~\bibnamefont {Hirano}}, \bibinfo {author} {\bibfnamefont
  {T.}~\bibnamefont {Mitsui}}, \bibinfo {author} {\bibfnamefont
  {H.}~\bibnamefont {Hosono}}, \ and\ \bibinfo {author} {\bibfnamefont
  {M.}~\bibnamefont {Seto}},\ }\href@noop {} {\bibfield  {journal} {\bibinfo
  {journal} {J. Phys. Soc. Japan}\ }\textbf {\bibinfo {volume} {77}},\ \bibinfo
  {pages} {103706} (\bibinfo {year} {2008})}\BibitemShut {NoStop}%
\bibitem [{\citenamefont {Nakai}\ \emph {et~al.}(2008)\citenamefont {Nakai},
  \citenamefont {Ishida}, \citenamefont {Kamihara}, \citenamefont {Hirano},\
  and\ \citenamefont {Hosono}}]{nakai}%
  \BibitemOpen
  \bibfield  {author} {\bibinfo {author} {\bibfnamefont {Y.}~\bibnamefont
  {Nakai}}, \bibinfo {author} {\bibfnamefont {K.}~\bibnamefont {Ishida}},
  \bibinfo {author} {\bibfnamefont {Y.}~\bibnamefont {Kamihara}}, \bibinfo
  {author} {\bibfnamefont {M.}~\bibnamefont {Hirano}}, \ and\ \bibinfo {author}
  {\bibfnamefont {H.}~\bibnamefont {Hosono}},\ }\href@noop {} {\bibfield
  {journal} {\bibinfo  {journal} {J. Phys. Soc. Japan}\ }\textbf {\bibinfo
  {volume} {77}},\ \bibinfo {pages} {073701} (\bibinfo {year}
  {2008})}\BibitemShut {NoStop}%
\bibitem [{\citenamefont {de~la Cruz}\ \emph {et~al.}(2008)\citenamefont {de~la
  Cruz}, \citenamefont {Huang}, \citenamefont {Lynn}, \citenamefont {Li},
  \citenamefont {II}, \citenamefont {Zarestky}, \citenamefont {Mook},
  \citenamefont {Chen}, \citenamefont {Luo}, \citenamefont {Wang},\ and\
  \citenamefont {Dai}}]{clarina}%
  \BibitemOpen
  \bibfield  {author} {\bibinfo {author} {\bibfnamefont {C.}~\bibnamefont
  {de~la Cruz}}, \bibinfo {author} {\bibfnamefont {Q.}~\bibnamefont {Huang}},
  \bibinfo {author} {\bibfnamefont {J.~W.}\ \bibnamefont {Lynn}}, \bibinfo
  {author} {\bibfnamefont {J.}~\bibnamefont {Li}}, \bibinfo {author}
  {\bibfnamefont {W.~R.}\ \bibnamefont {II}}, \bibinfo {author} {\bibfnamefont
  {J.~L.}\ \bibnamefont {Zarestky}}, \bibinfo {author} {\bibfnamefont {H.~A.}\
  \bibnamefont {Mook}}, \bibinfo {author} {\bibfnamefont {G.~F.}\ \bibnamefont
  {Chen}}, \bibinfo {author} {\bibfnamefont {J.~L.}\ \bibnamefont {Luo}},
  \bibinfo {author} {\bibfnamefont {N.~L.}\ \bibnamefont {Wang}}, \ and\
  \bibinfo {author} {\bibfnamefont {P.}~\bibnamefont {Dai}},\ }\href@noop {}
  {\bibfield  {journal} {\bibinfo  {journal} {nature}\ }\textbf {\bibinfo
  {volume} {453}},\ \bibinfo {pages} {899} (\bibinfo {year}
  {2008})}\BibitemShut {NoStop}%
\bibitem [{\citenamefont {Heisenberg}(1928)}]{hei}%
  \BibitemOpen
  \bibfield  {author} {\bibinfo {author} {\bibfnamefont {W.}~\bibnamefont
  {Heisenberg}},\ }\href@noop {} {\bibfield  {journal} {\bibinfo  {journal} {Z.
  Phys.}\ }\textbf {\bibinfo {volume} {49}},\ \bibinfo {pages} {619} (\bibinfo
  {year} {1928})}\BibitemShut {NoStop}%
\bibitem [{\citenamefont {Kr{\"u}ger}(2001)}]{enhm}%
  \BibitemOpen
  \bibfield  {author} {\bibinfo {author} {\bibfnamefont {E.}~\bibnamefont
  {Kr{\"u}ger}},\ }\href@noop {} {\bibfield  {journal} {\bibinfo  {journal}
  {Phys. Rev. B}\ }\textbf {\bibinfo {volume} {63}},\ \bibinfo {pages} {144403}
  (\bibinfo {year} {2001})}\BibitemShut {NoStop}%
\bibitem [{\citenamefont {Kr{\"u}ger}(2007)}]{ybacuo6}%
  \BibitemOpen
  \bibfield  {author} {\bibinfo {author} {\bibfnamefont {E.}~\bibnamefont
  {Kr{\"u}ger}},\ }\href@noop {} {\bibfield  {journal} {\bibinfo  {journal}
  {Phys. Rev. B}\ }\textbf {\bibinfo {volume} {75}},\ \bibinfo {pages} {024408}
  (\bibinfo {year} {2007})}\BibitemShut {NoStop}%
\bibitem [{\citenamefont {Kr{\"u}ger}(2010)}]{josybacuo7}%
  \BibitemOpen
  \bibfield  {author} {\bibinfo {author} {\bibfnamefont {E.}~\bibnamefont
  {Kr{\"u}ger}},\ }\href@noop {} {\bibfield  {journal} {\bibinfo  {journal} {J.
  Supercond.}\ }\textbf {\bibinfo {volume} {23}},\ \bibinfo {pages} {213}
  (\bibinfo {year} {2010})}\BibitemShut {NoStop}%
\bibitem [{\citenamefont {Blum}\ \emph {et~al.}(2009)\citenamefont {Blum},
  \citenamefont {Gehrke}, \citenamefont {Hanke}, \citenamefont {Havu},
  \citenamefont {Havu}, \citenamefont {Ren}, \citenamefont {Reuter},\ and\
  \citenamefont {Scheffler}}]{blum1}%
  \BibitemOpen
  \bibfield  {author} {\bibinfo {author} {\bibfnamefont {V.}~\bibnamefont
  {Blum}}, \bibinfo {author} {\bibfnamefont {R.}~\bibnamefont {Gehrke}},
  \bibinfo {author} {\bibfnamefont {F.}~\bibnamefont {Hanke}}, \bibinfo
  {author} {\bibfnamefont {P.}~\bibnamefont {Havu}}, \bibinfo {author}
  {\bibfnamefont {V.}~\bibnamefont {Havu}}, \bibinfo {author} {\bibfnamefont
  {X.}~\bibnamefont {Ren}}, \bibinfo {author} {\bibfnamefont {K.}~\bibnamefont
  {Reuter}}, \ and\ \bibinfo {author} {\bibfnamefont {M.}~\bibnamefont
  {Scheffler}},\ }\href@noop {} {\bibfield  {journal} {\bibinfo  {journal}
  {Computer Physics Communications}\ }\textbf {\bibinfo {volume} {180}},\
  \bibinfo {pages} {2175} (\bibinfo {year} {2009})}\BibitemShut {NoStop}%
\bibitem [{\citenamefont {Havu}\ \emph {et~al.}(2009)\citenamefont {Havu},
  \citenamefont {Blum}, \citenamefont {Havu},\ and\ \citenamefont
  {Scheffler}}]{blum2}%
  \BibitemOpen
  \bibfield  {author} {\bibinfo {author} {\bibfnamefont {V.}~\bibnamefont
  {Havu}}, \bibinfo {author} {\bibfnamefont {V.}~\bibnamefont {Blum}}, \bibinfo
  {author} {\bibfnamefont {P.}~\bibnamefont {Havu}}, \ and\ \bibinfo {author}
  {\bibfnamefont {M.}~\bibnamefont {Scheffler}},\ }\href@noop {} {\bibfield
  {journal} {\bibinfo  {journal} {Computer Physics Communications}\ }\textbf
  {\bibinfo {volume} {228}},\ \bibinfo {pages} {8367} (\bibinfo {year}
  {2009})}\BibitemShut {NoStop}%
\bibitem [{\citenamefont {Kamihara}\ \emph {et~al.}(2008)\citenamefont
  {Kamihara}, \citenamefont {Watanabe}, \citenamefont {Hirano},\ and\
  \citenamefont {Hosono}}]{kamihara}%
  \BibitemOpen
  \bibfield  {author} {\bibinfo {author} {\bibfnamefont {Y.}~\bibnamefont
  {Kamihara}}, \bibinfo {author} {\bibfnamefont {T.}~\bibnamefont {Watanabe}},
  \bibinfo {author} {\bibfnamefont {M.}~\bibnamefont {Hirano}}, \ and\ \bibinfo
  {author} {\bibfnamefont {H.}~\bibnamefont {Hosono}},\ }\href@noop {}
  {\bibfield  {journal} {\bibinfo  {journal} {J. Am. Chem. Soc.}\ }\textbf
  {\bibinfo {volume} {130}},\ \bibinfo {pages} {3296} (\bibinfo {year}
  {2008})}\BibitemShut {NoStop}%
\bibitem [{\citenamefont {Chen}\ \emph
  {et~al.}(2008{\natexlab{a}})\citenamefont {Chen}, \citenamefont {Li},
  \citenamefont {Li}, \citenamefont {Zhou}, \citenamefont {Wu}, \citenamefont
  {Dong}, \citenamefont {Hu}, \citenamefont {Zheng}, \citenamefont {Chen},
  \citenamefont {Yuan}, \citenamefont {Singleton}, \citenamefont {Luo},\ and\
  \citenamefont {Wang}}]{prl_chen}%
  \BibitemOpen
  \bibfield  {author} {\bibinfo {author} {\bibfnamefont {G.~F.}\ \bibnamefont
  {Chen}}, \bibinfo {author} {\bibfnamefont {Z.}~\bibnamefont {Li}}, \bibinfo
  {author} {\bibfnamefont {G.}~\bibnamefont {Li}}, \bibinfo {author}
  {\bibfnamefont {J.}~\bibnamefont {Zhou}}, \bibinfo {author} {\bibfnamefont
  {D.}~\bibnamefont {Wu}}, \bibinfo {author} {\bibfnamefont {J.}~\bibnamefont
  {Dong}}, \bibinfo {author} {\bibfnamefont {W.~Z.}\ \bibnamefont {Hu}},
  \bibinfo {author} {\bibfnamefont {P.}~\bibnamefont {Zheng}}, \bibinfo
  {author} {\bibfnamefont {Z.~J.}\ \bibnamefont {Chen}}, \bibinfo {author}
  {\bibfnamefont {H.~Q.}\ \bibnamefont {Yuan}}, \bibinfo {author}
  {\bibfnamefont {J.}~\bibnamefont {Singleton}}, \bibinfo {author}
  {\bibfnamefont {J.~L.}\ \bibnamefont {Luo}}, \ and\ \bibinfo {author}
  {\bibfnamefont {N.~L.}\ \bibnamefont {Wang}},\ }\href@noop {} {\bibfield
  {journal} {\bibinfo  {journal} {Phys. Rev. Lett.}\ }\textbf {\bibinfo
  {volume} {101}},\ \bibinfo {pages} {057007} (\bibinfo {year}
  {2008}{\natexlab{a}})}\BibitemShut {NoStop}%
\bibitem [{\citenamefont {Chen}\ \emph
  {et~al.}(2008{\natexlab{b}})\citenamefont {Chen}, \citenamefont {Wu},
  \citenamefont {Wu}, \citenamefont {Liu}, \citenamefont {Chen},\ and\
  \citenamefont {Fang}}]{nature_chen}%
  \BibitemOpen
  \bibfield  {author} {\bibinfo {author} {\bibfnamefont {X.~H.}\ \bibnamefont
  {Chen}}, \bibinfo {author} {\bibfnamefont {T.}~\bibnamefont {Wu}}, \bibinfo
  {author} {\bibfnamefont {G.}~\bibnamefont {Wu}}, \bibinfo {author}
  {\bibfnamefont {R.~H.}\ \bibnamefont {Liu}}, \bibinfo {author} {\bibfnamefont
  {H.}~\bibnamefont {Chen}}, \ and\ \bibinfo {author} {\bibfnamefont {D.~F.}\
  \bibnamefont {Fang}},\ }\href@noop {} {\bibfield  {journal} {\bibinfo
  {journal} {Nature}\ }\textbf {\bibinfo {volume} {453}},\ \bibinfo {pages}
  {761} (\bibinfo {year} {2008}{\natexlab{b}})}\BibitemShut {NoStop}%
\bibitem [{\citenamefont {Wen}\ \emph {et~al.}(2008)\citenamefont {Wen},
  \citenamefont {Mu}, \citenamefont {Fang}, \citenamefont {Yank},\ and\
  \citenamefont {Zhu}}]{wen}%
  \BibitemOpen
  \bibfield  {author} {\bibinfo {author} {\bibfnamefont {H.-H.}\ \bibnamefont
  {Wen}}, \bibinfo {author} {\bibfnamefont {G.}~\bibnamefont {Mu}}, \bibinfo
  {author} {\bibfnamefont {L.}~\bibnamefont {Fang}}, \bibinfo {author}
  {\bibfnamefont {H.}~\bibnamefont {Yank}}, \ and\ \bibinfo {author}
  {\bibfnamefont {X.}~\bibnamefont {Zhu}},\ }\href@noop {} {\bibfield
  {journal} {\bibinfo  {journal} {Europhys. Lett.}\ }\textbf {\bibinfo {volume}
  {82}},\ \bibinfo {pages} {17009} (\bibinfo {year} {2008})}\BibitemShut
  {NoStop}%
\bibitem [{\citenamefont {Dong}\ \emph {et~al.}(2008)\citenamefont {Dong},
  \citenamefont {Zhang}, \citenamefont {Xu}, \citenamefont {Li}, \citenamefont
  {Li}, \citenamefont {Hu}, \citenamefont {Wu}, \citenamefont {Chen},
  \citenamefont {Dai}, \citenamefont {Luo}, \citenamefont {Fang},\ and\
  \citenamefont {Wang}}]{dong}%
  \BibitemOpen
  \bibfield  {author} {\bibinfo {author} {\bibfnamefont {J.}~\bibnamefont
  {Dong}}, \bibinfo {author} {\bibfnamefont {H.~J.}\ \bibnamefont {Zhang}},
  \bibinfo {author} {\bibfnamefont {G.}~\bibnamefont {Xu}}, \bibinfo {author}
  {\bibfnamefont {Z.}~\bibnamefont {Li}}, \bibinfo {author} {\bibfnamefont
  {G.}~\bibnamefont {Li}}, \bibinfo {author} {\bibfnamefont {W.~Z.}\
  \bibnamefont {Hu}}, \bibinfo {author} {\bibfnamefont {D.}~\bibnamefont {Wu}},
  \bibinfo {author} {\bibfnamefont {G.~F.}\ \bibnamefont {Chen}}, \bibinfo
  {author} {\bibfnamefont {X.}~\bibnamefont {Dai}}, \bibinfo {author}
  {\bibfnamefont {J.~L.}\ \bibnamefont {Luo}}, \bibinfo {author} {\bibfnamefont
  {Z.}~\bibnamefont {Fang}}, \ and\ \bibinfo {author} {\bibfnamefont {N.~L.}\
  \bibnamefont {Wang}},\ }\href@noop {} {\bibfield  {journal} {\bibinfo
  {journal} {Europhys. Lett.}\ }\textbf {\bibinfo {volume} {83}},\ \bibinfo
  {pages} {27006} (\bibinfo {year} {2008})}\BibitemShut {NoStop}%
\bibitem [{\citenamefont {Bradley}\ and\ \citenamefont
  {A.P.Cracknell}(1972)}]{bc}%
  \BibitemOpen
  \bibfield  {author} {\bibinfo {author} {\bibfnamefont {C.}~\bibnamefont
  {Bradley}}\ and\ \bibinfo {author} {\bibnamefont {A.P.Cracknell}},\
  }\href@noop {} {\emph {\bibinfo {title} {The Mathematical Theory of Symmetry
  in Solids}}}\ (\bibinfo  {publisher} {Claredon, Oxford},\ \bibinfo {year}
  {1972})\BibitemShut {NoStop}%
\bibitem [{\citenamefont {Kr{\"u}ger}(1989)}]{ea}%
  \BibitemOpen
  \bibfield  {author} {\bibinfo {author} {\bibfnamefont {E.}~\bibnamefont
  {Kr{\"u}ger}},\ }\href@noop {} {\bibfield  {journal} {\bibinfo  {journal}
  {Phys. Rev. B}\ }\textbf {\bibinfo {volume} {40}},\ \bibinfo {pages} {11090}
  (\bibinfo {year} {1989})}\BibitemShut {NoStop}%
\bibitem [{\citenamefont {Kr{\"u}ger}(1999)}]{ef}%
  \BibitemOpen
  \bibfield  {author} {\bibinfo {author} {\bibfnamefont {E.}~\bibnamefont
  {Kr{\"u}ger}},\ }\href@noop {} {\bibfield  {journal} {\bibinfo  {journal}
  {Phys. Rev. B}\ }\textbf {\bibinfo {volume} {59}},\ \bibinfo {pages} {13795}
  (\bibinfo {year} {1999})}\BibitemShut {NoStop}%
\bibitem [{\citenamefont {Kr{\"u}ger}(2005)}]{josla2cuo4}%
  \BibitemOpen
  \bibfield  {author} {\bibinfo {author} {\bibfnamefont {E.}~\bibnamefont
  {Kr{\"u}ger}},\ }\href@noop {} {\bibfield  {journal} {\bibinfo  {journal} {J.
  Supercond.}\ }\textbf {\bibinfo {volume} {18(4)}},\ \bibinfo {pages} {433}
  (\bibinfo {year} {2005})}\BibitemShut {NoStop}%
\bibitem [{\citenamefont {Mott}(1956)}]{mott}%
  \BibitemOpen
  \bibfield  {author} {\bibinfo {author} {\bibfnamefont {N.~F.}\ \bibnamefont
  {Mott}},\ }\href@noop {} {\bibfield  {journal} {\bibinfo  {journal} {Can. J.
  Phys.}\ }\textbf {\bibinfo {volume} {34}},\ \bibinfo {pages} {1356} (\bibinfo
  {year} {1956})}\BibitemShut {NoStop}%
\bibitem [{\citenamefont {Hubbard}(1963)}]{hubbard}%
  \BibitemOpen
  \bibfield  {author} {\bibinfo {author} {\bibfnamefont {J.}~\bibnamefont
  {Hubbard}},\ }\href@noop {} {\bibfield  {journal} {\bibinfo  {journal} {Proc.
  R. Soc. London, Ser. A}\ }\textbf {\bibinfo {volume} {276}},\ \bibinfo
  {pages} {238} (\bibinfo {year} {1963})}\BibitemShut {NoStop}%
\bibitem [{\citenamefont {Yildirim}(2008)}]{yildirim}%
  \BibitemOpen
  \bibfield  {author} {\bibinfo {author} {\bibfnamefont {T.}~\bibnamefont
  {Yildirim}},\ }\href@noop {} {\bibfield  {journal} {\bibinfo  {journal}
  {Phys. Rev. Lett.}\ }\textbf {\bibinfo {volume} {101}},\ \bibinfo {pages}
  {057010} (\bibinfo {year} {2008})}\BibitemShut {NoStop}%
\bibitem [{\citenamefont {Kr{\"u}ger}(1985)}]{eabf}%
  \BibitemOpen
  \bibfield  {author} {\bibinfo {author} {\bibfnamefont {E.}~\bibnamefont
  {Kr{\"u}ger}},\ }\href@noop {} {\bibfield  {journal} {\bibinfo  {journal}
  {Phys. Rev. B}\ }\textbf {\bibinfo {volume} {32}},\ \bibinfo {pages} {7493}
  (\bibinfo {year} {1985})}\BibitemShut {NoStop}%
\end{thebibliography}

%

\appendix*
\FloatBarrier
\section{Tables}
\onecolumngrid
\FloatBarrier
\label{sec:tables}


\begin{table}[t]
\caption{
  Character tables of the single-valued irreducible representations of the
  space group $P4/nmm = \Gamma_qD^{7}_{4h}$ (129) of tetragonal
  paramagnetic LaFeAsO. 
\label{tab:rep129}}
\begin{tabular}[t]{ccccccccccc}
\multicolumn{11}{c}{$\Gamma (000)$, $Z (00\frac{1}{2})$}\\
 & $$ & $$ & $\{C^-_{4z}|0\frac{1}{2}0\}$ & $\{C_{2y}|0\frac{1}{2}0\}$ & $\{C_{2b}|000\}$ & $$ & $$ & $\{S^+_{4z}|0\frac{1}{2}0\}$ & $\{\sigma_y|0\frac{1}{2}0\}$ & $\{\sigma_{db}|000\}$\\
 & $\{E|000\}$ & $\{C_{2z}|\frac{1}{2}\frac{1}{2}0\}$ & $\{C^+_{4z}|\frac{1}{2}00\}$ & $\{C_{2x}|\frac{1}{2}00\}$ & $\{C_{2a}|\frac{1}{2}\frac{1}{2}0\}$ & $\{I|000\}$ & $\{\sigma_z|\frac{1}{2}\frac{1}{2}0\}$ & $\{S^-_{4z}|\frac{1}{2}00\}$ & $\{\sigma_x|\frac{1}{2}00\}$ & $\{\sigma_{da}|\frac{1}{2}\frac{1}{2}0\}$\\
\hline
$\Gamma^+_1$, $Z^+_1$ & 1 & 1 & 1 & 1 & 1 & 1 & 1 & 1 & 1 & 1\\
$\Gamma^+_2$, $Z^+_2$ & 1 & 1 & 1 & -1 & -1 & 1 & 1 & 1 & -1 & -1\\
$\Gamma^+_3$, $Z^+_3$ & 1 & 1 & -1 & 1 & -1 & 1 & 1 & -1 & 1 & -1\\
$\Gamma^+_4$, $Z^+_4$ & 1 & 1 & -1 & -1 & 1 & 1 & 1 & -1 & -1 & 1\\
$\Gamma^+_5$, $Z^+_5$ & 2 & -2 & 0 & 0 & 0 & 2 & -2 & 0 & 0 & 0\\
$\Gamma^-_1$, $Z^-_1$ & 1 & 1 & 1 & 1 & 1 & -1 & -1 & -1 & -1 & -1\\
$\Gamma^-_2$, $Z^-_2$ & 1 & 1 & 1 & -1 & -1 & -1 & -1 & -1 & 1 & 1\\
$\Gamma^-_3$, $Z^-_3$ & 1 & 1 & -1 & 1 & -1 & -1 & -1 & 1 & -1 & 1\\
$\Gamma^-_4$, $Z^-_4$ & 1 & 1 & -1 & -1 & 1 & -1 & -1 & 1 & 1 & -1\\
$\Gamma^-_5$, $Z^-_5$ & 2 & -2 & 0 & 0 & 0 & -2 & 2 & 0 & 0 & 0\\
\hline\\
\end{tabular}\hspace{1cm}
\begin{tabular}[t]{cccccccc}
\multicolumn{8}{c}{$M (\frac{1}{2}\frac{1}{2}0)$}\\
 & $$ & $$ & $$ & $$ & $$ & $$ & $$\\
 & $$ & $$ & $$ & $$ & $$ & $$ & $$\\
 & $$ & $$ & $$ & $$ & $\{\sigma_{da}|\frac{1}{2}\frac{1}{2}0\}$ & $\{\sigma_{da}|\frac{1}{2}\frac{3}{2}0\}$ & $\{I|010\}$\\
 & $\{E|000\}$ & $\{E|010\}$ & $\{C_{2z}|\frac{1}{2}\frac{1}{2}0\}$ & $\{C_{2z}|\frac{1}{2}\frac{3}{2}0\}$ & $\{\sigma_{db}|000\}$ & $\{\sigma_{db}|010\}$ & $\{I|000\}$\\
\hline
$M_1$ & 2 & -2 & 2 & -2 & 2 & -2 & 0\\
$M_2$ & 2 & -2 & -2 & 2 & 0 & 0 & 0\\
$M_3$ & 2 & -2 & 2 & -2 & -2 & 2 & 0\\
$M_4$ & 2 & -2 & -2 & 2 & 0 & 0 & 0\\
\hline\\
\end{tabular}\hspace{1cm}
\begin{tabular}[t]{cccccccc}
\multicolumn{8}{c}{$M (\frac{1}{2}\frac{1}{2}0)$\qquad $(continued)$}\\
 & $$ & $$ & $$ & $\{\sigma_y|0\frac{3}{2}0\}$ & $\{C^+_{4z}|\frac{1}{2}10\}$ & $\{C_{2y}|0\frac{3}{2}0\}$ & $\{S^-_{4z}|\frac{1}{2}10\}$\\
 & $$ & $$ & $$ & $\{\sigma_y|0\frac{1}{2}0\}$ & $\{C^-_{4z}|0\frac{3}{2}0\}$ & $\{C_{2y}|0\frac{1}{2}0\}$ & $\{S^+_{4z}|0\frac{1}{2}0\}$\\
 & $\{\sigma_z|\frac{1}{2}\frac{1}{2}0\}$ & $\{C_{2b}|000\}$ & $\{C_{2b}|010\}$ & $\{\sigma_x|\frac{1}{2}10\}$ & $\{C^-_{4z}|0\frac{1}{2}0\}$ & $\{C_{2x}|\frac{1}{2}00\}$ & $\{S^+_{4z}|0\frac{3}{2}0\}$\\
 & $\{\sigma_z|\frac{1}{2}\frac{3}{2}0\}$ & $\{C_{2a}|\frac{1}{2}\frac{3}{2}0\}$ & $\{C_{2a}|\frac{1}{2}\frac{1}{2}0\}$ & $\{\sigma_x|\frac{1}{2}00\}$ & $\{C^+_{4z}|\frac{1}{2}00\}$ & $\{C_{2x}|\frac{1}{2}10\}$ & $\{S^-_{4z}|\frac{1}{2}00\}$\\
\hline
$M_1$ & 0 & 0 & 0 & 0 & 0 & 0 & 0\\
$M_2$ & 0 & 2 & -2 & 0 & 0 & 0 & 0\\
$M_3$ & 0 & 0 & 0 & 0 & 0 & 0 & 0\\
$M_4$ & 0 & -2 & 2 & 0 & 0 & 0 & 0\\
\hline\\
\end{tabular}\hspace{1cm}
\end{table}
\begin{table}[t]
\begin{tabular}{cccccccc}
\multicolumn{8}{c}{$A (\frac{1}{2}\frac{1}{2}\frac{1}{2})$}\\
 & $$ & $$ & $$ & $$ & $$ & $$ & $$\\
 & $$ & $$ & $$ & $$ & $$ & $$ & $$\\
 & $$ & $$ & $$ & $$ & $\{\sigma_{da}|\frac{1}{2}\frac{1}{2}0\}$ & $\{\sigma_{da}|\frac{1}{2}\frac{1}{2}1\}$ & $\{I|001\}$\\
 & $\{E|000\}$ & $\{E|001\}$ & $\{C_{2z}|\frac{1}{2}\frac{1}{2}0\}$ & $\{C_{2z}|\frac{1}{2}\frac{1}{2}1\}$ & $\{\sigma_{db}|000\}$ & $\{\sigma_{db}|001\}$ & $\{I|000\}$\\
\hline
$A_1$ & 2 & -2 & 2 & -2 & 2 & -2 & 0\\
$A_2$ & 2 & -2 & -2 & 2 & 0 & 0 & 0\\
$A_3$ & 2 & -2 & 2 & -2 & -2 & 2 & 0\\
$A_4$ & 2 & -2 & -2 & 2 & 0 & 0 & 0\\
\hline\\
\end{tabular}\hspace{1cm}
\begin{tabular}[t]{cccccccc}
\multicolumn{8}{c}{$A (\frac{1}{2}\frac{1}{2}\frac{1}{2})$\qquad $(continued)$}\\
 & $$ & $$ & $$ & $\{\sigma_y|0\frac{1}{2}1\}$ & $\{C^+_{4z}|\frac{1}{2}01\}$ & $\{C_{2y}|0\frac{1}{2}1\}$ & $\{S^-_{4z}|\frac{1}{2}01\}$\\
 & $$ & $$ & $$ & $\{\sigma_y|0\frac{1}{2}0\}$ & $\{C^-_{4z}|0\frac{1}{2}1\}$ & $\{C_{2y}|0\frac{1}{2}0\}$ & $\{S^+_{4z}|0\frac{1}{2}0\}$\\
 & $\{\sigma_z|\frac{1}{2}\frac{1}{2}0\}$ & $\{C_{2b}|000\}$ & $\{C_{2b}|001\}$ & $\{\sigma_x|\frac{1}{2}01\}$ & $\{C^-_{4z}|0\frac{1}{2}0\}$ & $\{C_{2x}|\frac{1}{2}00\}$ & $\{S^+_{4z}|0\frac{1}{2}1\}$\\
 & $\{\sigma_z|\frac{1}{2}\frac{1}{2}1\}$ & $\{C_{2a}|\frac{1}{2}\frac{1}{2}1\}$ & $\{C_{2a}|\frac{1}{2}\frac{1}{2}0\}$ & $\{\sigma_x|\frac{1}{2}00\}$ & $\{C^+_{4z}|\frac{1}{2}00\}$ & $\{C_{2x}|\frac{1}{2}01\}$ & $\{S^-_{4z}|\frac{1}{2}00\}$\\
\hline
$A_1$ & 0 & 0 & 0 & 0 & 0 & 0 & 0\\
$A_2$ & 0 & 2 & -2 & 0 & 0 & 0 & 0\\
$A_3$ & 0 & 0 & 0 & 0 & 0 & 0 & 0\\
$A_4$ & 0 & -2 & 2 & 0 & 0 & 0 & 0\\
\hline\\
\end{tabular}\hspace{1cm}
\begin{tabular}[t]{ccccccccccc}
\multicolumn{11}{c}{$R (0\frac{1}{2}\frac{1}{2})$}\\
 & $$ & $$ & $\{C_{2y}|0\frac{1}{2}1\}$ & $\{C_{2x}|\frac{1}{2}00\}$ & $\{C_{2z}|\frac{1}{2}\frac{1}{2}0\}$ & $$ & $$ & $\{\sigma_z|\frac{1}{2}\frac{1}{2}1\}$ & $\{I|000\}$ & $\{\sigma_y|0\frac{1}{2}0\}$\\
 & $\{E|000\}$ & $\{E|001\}$ & $\{C_{2y}|0\frac{1}{2}0\}$ & $\{C_{2x}|\frac{1}{2}01\}$ & $\{C_{2z}|\frac{1}{2}\frac{1}{2}1\}$ & $\{\sigma_x|\frac{1}{2}00\}$ & $\{\sigma_x|\frac{1}{2}01\}$ & $\{\sigma_z|\frac{1}{2}\frac{1}{2}0\}$ & $\{I|001\}$ & $\{\sigma_y|0\frac{1}{2}1\}$\\
\hline
$R_1$ & 2 & -2 & 0 & 0 & 0 & 2 & -2 & 0 & 0 & 0\\
$R_2$ & 2 & -2 & 0 & 0 & 0 & -2 & 2 & 0 & 0 & 0\\
\hline\\
\end{tabular}\hspace{1cm}
\begin{tabular}[t]{ccccccccccc}
\multicolumn{11}{c}{$X (0\frac{1}{2}0)$}\\
 & $$ & $$ & $\{C_{2y}|0\frac{3}{2}0\}$ & $\{C_{2x}|\frac{1}{2}00\}$ & $\{C_{2z}|\frac{1}{2}\frac{1}{2}0\}$ & $$ & $$ & $\{\sigma_z|\frac{1}{2}\frac{3}{2}0\}$ & $\{I|000\}$ & $\{\sigma_y|0\frac{1}{2}0\}$\\
 & $\{E|000\}$ & $\{E|010\}$ & $\{C_{2y}|0\frac{1}{2}0\}$ & $\{C_{2x}|\frac{1}{2}10\}$ & $\{C_{2z}|\frac{1}{2}\frac{3}{2}0\}$ & $\{\sigma_x|\frac{1}{2}00\}$ & $\{\sigma_x|\frac{1}{2}10\}$ & $\{\sigma_z|\frac{1}{2}\frac{1}{2}0\}$ & $\{I|010\}$ & $\{\sigma_y|0\frac{3}{2}0\}$\\
\hline
$X_1$ & 2 & -2 & 0 & 0 & 0 & 2 & -2 & 0 & 0 & 0\\
$X_2$ & 2 & -2 & 0 & 0 & 0 & -2 & 2 & 0 & 0 & 0\\
\hline\\
\end{tabular}\hspace{1cm}
\ \\
\begin{flushleft}
Notes to Table~\ref{tab:rep129}
\end{flushleft}
\begin{enumerate}
\item   The space group operations are related to the coordinate systems in 
  Figs.~\ref{fig:cartesian} and~\ref{fig:structures} (a).
\item The character table is determined from Table 5.7 of
  Ref.~\protect\cite{bc}.  The origin of the coordinate system used in
  Ref.~\protect\cite{bc} is translated into the origin used in this paper by
  $\bm t_0 = -\frac{1}{4}\bm T_1 - \frac{1}{4}\bm T_2$. That is, the symmetry
  operations $P_{bc}$ given in Table 5.7 of Ref.~\protect\cite{bc} are changed
  into the operations $P_{p}$ used in this paper by the equation  
  $$P_{p} = \{E|\bm t_0\}P_{bc}\{E|-\bm t_0\}, $$
where $E$ is the identity operation, cf. Eq.~(3.5.11) of Ref.~\cite{bc} (which
is related to the opposite translation $-\bm t_0$).
\end{enumerate}
\end{table}


\FloatBarrier


\begin{table}[t]
\caption{
  Character tables of the single-valued irreducible representations of the
  orthorhombic space group $Imma = \Gamma_o^vD^{28}_{2h}$ (74) of the antiferromagnetic
  structure depicted in Fig.~\ref{fig:structures} (b).  
  \label{tab:rep74}}
\begin{tabular}[t]{cccccccccc}
\multicolumn{10}{c}{$\Gamma (000)$}\\
 & $K$ & $\{E|000\}$ & $\{C_{2z}|0\frac{1}{2}\frac{1}{2}\}$ & $\{C_{2a}|\frac{1}{2}0\frac{1}{2}\}$ & $\{C_{2b}|\frac{1}{2}\frac{1}{2}0\}$ & $\{I|\frac{1}{2}\frac{1}{2}0\}$ & $\{\sigma_z|\frac{1}{2}0\frac{1}{2}\}$ & $\{\sigma_{da}|0\frac{1}{2}\frac{1}{2}\}$ & $\{\sigma_{db}|000\}$\\
\hline
$\Gamma^+_1$ & (a) & 1 & 1 & 1 & 1 & 1 & 1 & 1 & 1\\
$\Gamma^+_2$ & (a) & 1 & -1 & 1 & -1 & 1 & -1 & 1 & -1\\
$\Gamma^+_3$ & (a) & 1 & 1 & -1 & -1 & 1 & 1 & -1 & -1\\
$\Gamma^+_4$ & (a) & 1 & -1 & -1 & 1 & 1 & -1 & -1 & 1\\
$\Gamma^-_1$ & (a) & 1 & 1 & 1 & 1 & -1 & -1 & -1 & -1\\
$\Gamma^-_2$ & (a) & 1 & -1 & 1 & -1 & -1 & 1 & -1 & 1\\
$\Gamma^-_3$ & (a) & 1 & 1 & -1 & -1 & -1 & -1 & 1 & 1\\
$\Gamma^-_4$ & (a) & 1 & -1 & -1 & 1 & -1 & 1 & 1 & -1\\
\hline\\
\end{tabular}\hspace{1cm}
\begin{tabular}[t]{cccccccccc}
\multicolumn{10}{c}{$X (\frac{1}{2}\frac{1}{2}\overline{\frac{1}{2}})$}\\
 & $K$ & $\{E|000\}$ & $\{C_{2z}|0\frac{1}{2}\frac{1}{2}\}$ & $\{C_{2a}|\frac{1}{2}0\frac{1}{2}\}$ & $\{C_{2b}|\frac{1}{2}\frac{1}{2}1\}$ & $\{I|\frac{1}{2}\frac{1}{2}1\}$ & $\{\sigma_z|\frac{1}{2}0\frac{1}{2}\}$ & $\{\sigma_{da}|0\frac{1}{2}\frac{1}{2}\}$ & $\{\sigma_{db}|000\}$\\
\hline
$X^+_1$ & (a) & 1 & 1 & 1 & 1 & 1 & 1 & 1 & 1\\
$X^+_2$ & (a) & 1 & -1 & 1 & -1 & 1 & -1 & 1 & -1\\
$X^+_3$ & (a) & 1 & 1 & -1 & -1 & 1 & 1 & -1 & -1\\
$X^+_4$ & (a) & 1 & -1 & -1 & 1 & 1 & -1 & -1 & 1\\
$X^-_1$ & (a) & 1 & 1 & 1 & 1 & -1 & -1 & -1 & -1\\
$X^-_2$ & (a) & 1 & -1 & 1 & -1 & -1 & 1 & -1 & 1\\
$X^-_3$ & (a) & 1 & 1 & -1 & -1 & -1 & -1 & 1 & 1\\
$X^-_4$ & (a) & 1 & -1 & -1 & 1 & -1 & 1 & 1 & -1\\
\hline\\
\end{tabular}\hspace{1cm}
\begin{tabular}[t]{cccccc}
\multicolumn{6}{c}{$R (\frac{1}{2}00)$}\\
 & $K$ & $\{E|000\}$ & $\{C_{2a}|\frac{1}{2}0\frac{1}{2}\}$ & $\{I|\frac{1}{2}\frac{1}{2}0\}$ & $\{\sigma_{da}|0\frac{1}{2}\frac{1}{2}\}$\\
\hline
$R^+_1$ & (a) & 1 & 1 & 1 & 1\\
$R^-_1$ & (a) & 1 & 1 & -1 & -1\\
$R^+_2$ & (a) & 1 & -1 & 1 & -1\\
$R^-_2$ & (a) & 1 & -1 & -1 & 1\\
\hline\\
\end{tabular}\hspace{1cm}
\begin{tabular}[t]{cccccc}
\multicolumn{6}{c}{$S (\frac{1}{2}0\overline{\frac{1}{2}})$}\\
 & $K$ & $\{E|000\}$ & $\{C_{2b}|\frac{1}{2}\frac{1}{2}0\}$ & $\{I|\frac{1}{2}\frac{1}{2}1\}$ & $\{\sigma_{db}|001\}$\\
\hline
$S^+_1$ & (a) & 1 & 1 & 1 & 1\\
$S^-_1$ & (a) & 1 & 1 & -1 & -1\\
$S^+_2$ & (a) & 1 & -1 & 1 & -1\\
$S^-_2$ & (a) & 1 & -1 & -1 & 1\\
\hline\\
\end{tabular}\hspace{1cm}
\ \\
\begin{flushleft}
Notes to Table~\ref{tab:rep74}
\end{flushleft}
\begin{enumerate}
\item The points $T$ and $W$ are not listed 
  because they possess only one two-dimensional representation $T_1$ and
  $W_1$, respectively. 
\item The character tables are determined from Table 5.7 in
  Ref.~\protect\cite{bc}. The origin of the coordinate system used in
  Ref.~\protect\cite{bc} is translated into the origin used in this paper by
  $\bm t_0 = \frac{1}{2}\bm T_3$. That is, the symmetry
  operations $P_{bc}$ given in Table 5.7 of Ref.~\protect\cite{bc} are changed
  into the operations $P_{p}$ used in this paper by the equation  
  $$P_{p} = \{E|\bm t_0\}P_{bc}\{E|-\bm t_0\}, $$
  where $E$ is the identity operation, cf. Eq.~(3.5.11) of
  Ref.~\cite{bc} (which is related to the opposite translation $-\bm
  t_0$).
\item The space group operations are related to the coordinate systems in
  Figs.~\ref{fig:cartesian} and ~\ref{fig:structures} (b). The $x$, $y$, and
  $z$ axes have the same orientations as in the tetragonal structure in
  Fig.~\ref{fig:structures} (a). In this way, $Imma$ becomes a subgroup of the
  tetragonal group $P4/nmm$. As a consequence, the $x y z$ coordinate system
  used in Ref.~\protect\cite{bc} is rotated anti-clockwise with respect to the
  basic translations through $\frac{\pi}{4}$ radians about the $z$ axis. Thus,
  the $x$ and $y$ axes in Ref.~\protect\cite{bc} become the $B$ and $A$ axes,
  respectively, in this paper, see Fig.~\ref{fig:cartesian}. Consequently, the
  point group operations belonging to the space group operations $P_{p}$
  calculated by the above equation are renamed in a second step.  For
  instance, the operation $C_{2x}$ in Table 5.7 of Ref.~\protect\cite{bc}
  turns into the operation $C_{2b}$.
\item $K$ stands for the operator of time-inversion. The entries below $K$
  specify whether the related representation follows, with respect to the
  magnetic group $Imma + KImma$, case $(a)$, case $(b)$, or case $(c)$ when
  they are given by Eqs.~(7.3.45-47) of Ref.~\protect\cite{bc}.
\item The cases $(a)$, $(b)$, and $(c)$ are determined by Eq.~\gl{eq:18}.
\item All the one-dimensional representations of the space group $Imma$ are
  real and, hence, follow case (a) with respect to $Imma + KImma$.  Therefore,
  stable magnetic structures with this space group do not exist, see
  Theorem~\ref{theorem}.
\end{enumerate}
\end{table}


\FloatBarrier


\begin{table}[!]
\caption{
  Character tables of the single-valued irreducible representations of the 
  orthorhombic space group $Pnn2 = \Gamma_oC^{10}_{2v}$ (34) of the 
  antiferromagnetic structure depicted in Fig.~\ref{fig:structures} (c).  
  \label{tab:rep34}}
\begin{tabular}[t]{cccccccc}
\multicolumn{8}{c}{$\Gamma (000)$}\\
 & $K$ & $\{K|\frac{1}{2}\frac{1}{2}0\}$ & $\{K|00\frac{1}{2}\}$ & $\{E|000\}$ & $\{C_{2z}|\frac{1}{2}00\}$ & $\{\sigma_{da}|0\frac{1}{2}\frac{1}{2}\}$ & $\{\sigma_{db}|\frac{1}{2}\frac{1}{2}\frac{1}{2}\}$\\
\hline
$\Gamma_1$ & (a) & (a) & (a) & 1 & 1 & 1 & 1\\
$\Gamma_3$ & (a) & (a) & (a) & 1 & 1 & -1 & -1\\
$\Gamma_2$ & (a) & (a) & (a) & 1 & -1 & 1 & -1\\
$\Gamma_4$ & (a) & (a) & (a) & 1 & -1 & -1 & 1\\
\hline\\
\end{tabular}\hspace{1cm}
\begin{tabular}[t]{cccccccccccc}
\multicolumn{12}{c}{$Z (00\frac{1}{2})$}\\
 & $K$ & $\{K|\frac{1}{2}\frac{1}{2}0\}$ & $\{K|00\frac{1}{2}\}$ & $\{E|000\}$ & $\{\sigma_{da}|0\frac{1}{2}\frac{1}{2}\}$ & $\{E|001\}$ & $\{\sigma_{da}|0\frac{1}{2}\frac{3}{2}\}$ & $\{C_{2z}|\frac{1}{2}00\}$ & $\{\sigma_{db}|\frac{1}{2}\frac{1}{2}\frac{1}{2}\}$ & $\{C_{2z}|\frac{1}{2}01\}$ & $\{\sigma_{db}|\frac{1}{2}\frac{1}{2}\frac{3}{2}\}$\\
\hline
$Z_1$ & (c) & (c) & (c) & 1 & i & -1 & -i & 1 & i & -1 & -i\\
$Z_2$ & (c) & (c) & (c) & 1 & -i & -1 & i & 1 & -i & -1 & i\\
$Z_3$ & (c) & (c) & (c) & 1 & i & -1 & -i & -1 & -i & 1 & i\\
$Z_4$ & (c) & (c) & (c) & 1 & -i & -1 & i & -1 & i & 1 & -i\\
\hline\\
\end{tabular}\hspace{1cm}
\begin{tabular}[t]{cccccccccccc}
\multicolumn{12}{c}{$S (\overline{\frac{1}{2}}\frac{1}{2}0)$}\\
 & $K$ & $\{K|\frac{1}{2}\frac{1}{2}0\}$ & $\{K|00\frac{1}{2}\}$ & $\{E|000\}$ & $\{\sigma_{da}|0\frac{1}{2}\frac{1}{2}\}$ & $\{E|010\}$ & $\{\sigma_{da}|0\frac{3}{2}\frac{1}{2}\}$ & $\{C_{2z}|\frac{1}{2}10\}$ & $\{\sigma_{db}|\frac{1}{2}\frac{1}{2}\frac{1}{2}\}$ & $\{C_{2z}|\frac{1}{2}00\}$ & $\{\sigma_{db}|\frac{1}{2}\frac{3}{2}\frac{1}{2}\}$\\
\hline
$S_1$ & (c) & (a) & (c) & 1 & i & -1 & -i & 1 & i & -1 & -i\\
$S_2$ & (c) & (a) & (c) & 1 & -i & -1 & i & 1 & -i & -1 & i\\
$S_3$ & (c) & (a) & (c) & 1 & i & -1 & -i & -1 & -i & 1 & i\\
$S_4$ & (c) & (a) & (c) & 1 & -i & -1 & i & -1 & i & 1 & -i\\
\hline\\
\end{tabular}\hspace{1cm}
\ \\
\begin{flushleft}
Notes to Table~\ref{tab:rep34}
\end{flushleft}
\begin{enumerate}
\item In addition to $\Gamma$ only the points $Z$ and $S$ are listed because
  only $Z$ and $S$ possess one-dimensional non-real representations.
\item The character tables are determined from Table 5.7 in
  Ref.~\protect\cite{bc}.  The origin of the coordinate system used in
  Ref.~\protect\cite{bc} is translated into the origin used in this paper by
  $\bm t_0 = \frac{1}{4}\bm T_2$. That is, the symmetry
  operations $P_{bc}$ given in Table 5.7 of Ref.~\protect\cite{bc} are changed
  into the operations $P_{p}$ used in this paper by the equation  
  $$P_{p} = \{E|\bm t_0\}P_{bc}\{E|-\bm t_0\}, $$
  where $E$ is the identity operation, cf. Eq.~(3.5.11) of Ref.~\cite{bc}
  (which is related to the opposite translation $-\bm t_0$).
\item The space group operations are related to the coordinate systems in
  Figs.~\ref{fig:cartesian} and~\ref{fig:structures} (c). The $x$, $y$, and
  $z$ axes have the same orientations as in the tetragonal structure in
  Fig.~\ref{fig:structures} (a). In this way, $Pnn2$ becomes a subgroup of the
  tetragonal group $P4/nmm$. As a consequence, the $x y z$ coordinate system
  used in Ref.~\protect\cite{bc} is rotated anti-clockwise with respect to the
  basic translations through $\frac{\pi}{4}$ radians about the $z$ axis. Thus,
  the $x$ and $y$ axes in Ref.~\protect\cite{bc} become the $B$ and $A$ axes,
  respectively, in this paper, see Fig.~\ref{fig:cartesian}. Consequently, the
  point group operations belonging to the space group operations $P_{p}$
  calculated by the above equation are renamed in a second step. For instance,
  the operation $\sigma_{x}$ in Table 5.7 of Ref.~\protect\cite{bc} turns into
  the operation $\sigma_{db}$.
\item $K$ stands for the operator of time inversion. The entries below $K$ and
  the operators $\{K|\bm t\}$ specify whether the related representation
  follows, with respect to the magnetic groups $Pnn2 + KPnn2$ and $Pnn2 +
  \{K|\bm t\}Pnn2$, respectively, case $(a)$, case $(b)$, or case $(c)$ when
  they are given by Eqs.~(7.3.45-47) of Ref.~\protect\cite{bc}.
\item The cases $(a)$, $(b)$, and $(c)$ are determined by Eq.~\gl{eq:18}.
\item The little group of point $S$ comprises the whole space group $Pnn2$ and
  $S$ possesses one-dimensional representations following case $(c)$ and case
  $(a)$ with respect to $Pnn2 + KPnn2$ and $Pnn2 +
  \{K|\frac{1}{2}\frac{1}{2}0\}Pnn2$, respectively. Consequently, a magnetic
  structure with the space group $Pnn2$ can be stable, see
  Theorem~\ref{theorem}.
\end{enumerate}
\end{table}


\FloatBarrier

\begin{table}[!]
\caption{
Compatibility relations between the Brillouin zone for tetragonal
paramagnetic LaFeAsO and the Brillouin zone for the orthorhombic
antiferromagnetic structure depicted in Fig.~\ref{fig:structures} (b).  
\label{tab:falten}
}
\begin{tabular}[t]{cccccccccc}
\multicolumn{10}{c}{$\Gamma$}\\
\hline
$\Gamma^+_1$ & $\Gamma^+_2$ & $\Gamma^+_3$ & $\Gamma^+_4$ & $\Gamma^+_5$ & $\Gamma^-_1$ & $\Gamma^-_2$ & $\Gamma^-_3$ & $\Gamma^-_4$ & $\Gamma^-_5$\\
$\Gamma^+_1$ & $\Gamma^+_3$ & $\Gamma^+_3$ & $\Gamma^+_1$ & $\Gamma^+_2$ + $\Gamma^+_4$ & $\Gamma^-_1$ & $\Gamma^-_3$ & $\Gamma^-_3$ & $\Gamma^-_1$ & $\Gamma^-_2$ + $\Gamma^-_4$\\
\hline\\
\end{tabular}\hspace{1cm}
\begin{tabular}[t]{cccc}
\multicolumn{4}{c}{$X$}\\
\hline
$M_1$ & $M_2$ & $M_3$ & $M_4$\\
$X^+_1$ + $X^-_3$ & $X^+_2$ + $X^-_2$ & $X^+_3$ + $X^-_1$ & $X^+_4$ + $X^-_4$\\
\hline\\
\end{tabular}\hspace{1cm}
\begin{tabular}[t]{cccccccccc}
\multicolumn{10}{c}{$X$}\\
\hline
$Z^+_1$ & $Z^+_2$ & $Z^+_3$ & $Z^+_4$ & $Z^+_5$ & $Z^-_1$ & $Z^-_2$ & $Z^-_3$ & $Z^-_4$ & $Z^-_5$\\
$X^+_1$ & $X^+_3$ & $X^+_3$ & $X^+_1$ & $X^+_2$ + $X^+_4$ & $X^-_1$ & $X^-_3$ & $X^-_3$ & $X^-_1$ & $X^-_2$ + $X^-_4$\\
\hline\\
\end{tabular}\hspace{1cm}
\begin{tabular}[t]{cccc}
\multicolumn{4}{c}{$\Gamma$}\\
\hline
$A_1$ & $A_2$ & $A_3$ & $A_4$\\
$\Gamma^+_1$ + $\Gamma^-_3$ & $\Gamma^+_2$ + $\Gamma^-_2$ & $\Gamma^+_3$ + $\Gamma^-_1$ & $\Gamma^+_4$ + $\Gamma^-_4$\\
\hline\\
\end{tabular}\hspace{5cm}
\begin{tabular}[t]{cc}
\multicolumn{2}{c}{$T$}\\
\hline
$R_1$ & $R_2$\\
$T_1$ & $T_1$\\
\hline\\
\end{tabular}\hspace{1cm}
\begin{tabular}[t]{cc}
\multicolumn{2}{c}{$T$}\\
\hline
$X_1$ & $X_2$\\
$T_1$ & $T_1$\\
\hline\\
\end{tabular}\hspace{1cm}
\ \\
\begin{flushleft}
Notes to Table~\ref{tab:falten}
\end{flushleft}
\begin{enumerate}
\item The antiferromagnetic structure in Fig.~\ref{fig:structures} (b) has the
  space group $Imma$.
\item The Brillouin zone for the orthorhombic space group $Imma$ lies
  within the Brillouin zone for the tetragonal space group $P4/nmm$.
  The points $\Gamma$ and $A$ in the tetragonal Brillouin zone are
  equivalent to the point $\Gamma$ in the orthorhombic Brillouin zone; $M$
  and $Z$ are equivalent to $X$, and $R$ and $X$ are equivalent to $T$.
\item The upper rows list the representations of the little groups of the
  points of symmetry in the Brillouin zone for the tetragonal paramagnetic
  phase. The lower rows list
  representations of the little groups of the related points of symmetry in
  the Brillouin zone for the antiferromagnetic structure.
  
  The representations in the same column are compatible in the
  following sense: Bloch functions that are basis functions of a
  representation $D_i$ in the upper row can be unitarily transformed into
  the basis functions of the representation given below $D_i$.
\item The compatibility relations are determined in the way described in
  great detail in Ref.~\cite{eabf}.
\item The representations are labeled as given in Tables~\ref{tab:rep129} and 
 \ref{tab:rep74}, respectively.
\end{enumerate}
\end{table}


\FloatBarrier

\begin{table}
\caption{
Single-valued representations of all the magnetic bands related to the space
group $Imma = \Gamma_o^vD^{28}_{2h}$ (74) with the Wannier functions being
centered (a) at the La atoms, (b) at the Fe atoms, (c) at the As atoms, and
(d) at the O atoms.    
\label{tab:wf74}}
\begin{tabular}[t]{ccccccc}
\multicolumn{7}{c}{(a)\ \ La}\\
 & $\Gamma$ & $X$ & $R$ & $S$ & $T$ & $W$\\
\hline
Band 1\ \ & 2$\Gamma^+_1$ + 2$\Gamma^-_3$ & 2$X^+_1$ + 2$X^-_3$ & 2$R^+_1$ + 2$R^-_2$ & 2$S^-_1$ + 2$S^+_2$ & 2$T_1$ & 2$W_1$\\
Band 2\ \ & 2$\Gamma^+_2$ + 2$\Gamma^-_4$ & 2$X^+_2$ + 2$X^-_4$ & 2$R^+_1$ + 2$R^-_2$ & 2$S^+_1$ + 2$S^-_2$ & 2$T_1$ & 2$W_1$\\
Band 3\ \ & 2$\Gamma^+_3$ + 2$\Gamma^-_1$ & 2$X^+_3$ + 2$X^-_1$ & 2$R^-_1$ + 2$R^+_2$ & 2$S^+_1$ + 2$S^-_2$ & 2$T_1$ & 2$W_1$\\
Band 4\ \ & 2$\Gamma^+_4$ + 2$\Gamma^-_2$ & 2$X^+_4$ + 2$X^-_2$ & 2$R^-_1$ + 2$R^+_2$ & 2$S^-_1$ + 2$S^+_2$ & 2$T_1$ & 2$W_1$\\
\hline\\
\end{tabular}
\begin{tabular}[t]{ccccccc}
\multicolumn{7}{c}{(b)\ \ Fe}\\
 & $\Gamma$ & $X$ & $R$ & $S$ & $T$ & $W$\\
\hline
Band 1\ \ & $\Gamma^+_1$ + $\Gamma^+_4$ + $\Gamma^-_1$ + $\Gamma^-_4$ & $X^+_2$ + $X^+_3$ + $X^-_2$ + $X^-_3$ & $R^+_1$ + $R^-_1$ + $R^+_2$ + $R^-_2$ & $S^+_1$ + $S^-_1$ + $S^+_2$ + $S^-_2$ & 2$T_1$ & 2$W_1$\\
Band 2\ \ & $\Gamma^+_2$ + $\Gamma^+_3$ + $\Gamma^-_2$ + $\Gamma^-_3$ & $X^+_1$ + $X^+_4$ + $X^-_1$ + $X^-_4$ & $R^+_1$ + $R^-_1$ + $R^+_2$ + $R^-_2$ & $S^+_1$ + $S^-_1$ + $S^+_2$ + $S^-_2$ & 2$T_1$ & 2$W_1$\\
\hline\\
\end{tabular}
\begin{tabular}[t]{ccccccc}
\multicolumn{7}{c}{(c)\ \ As}\\
 & $\Gamma$ & $X$ & $R$ & $S$ & $T$ & $W$\\
\hline
Band 1\ \ & 2$\Gamma^+_1$ + 2$\Gamma^-_3$ & 2$X^+_1$ + 2$X^-_3$ & 2$R^+_1$ + 2$R^-_2$ & 2$S^-_1$ + 2$S^+_2$ & 2$T_1$ & 2$W_1$\\
Band 2\ \ & 2$\Gamma^+_2$ + 2$\Gamma^-_4$ & 2$X^+_2$ + 2$X^-_4$ & 2$R^+_1$ + 2$R^-_2$ & 2$S^+_1$ + 2$S^-_2$ & 2$T_1$ & 2$W_1$\\
Band 3\ \ & 2$\Gamma^+_3$ + 2$\Gamma^-_1$ & 2$X^+_3$ + 2$X^-_1$ & 2$R^-_1$ + 2$R^+_2$ & 2$S^+_1$ + 2$S^-_2$ & 2$T_1$ & 2$W_1$\\
Band 4\ \ & 2$\Gamma^+_4$ + 2$\Gamma^-_2$ & 2$X^+_4$ + 2$X^-_2$ & 2$R^-_1$ + 2$R^+_2$ & 2$S^-_1$ + 2$S^+_2$ & 2$T_1$ & 2$W_1$\\
\hline\\
\end{tabular}
\begin{tabular}[t]{ccccccc}
\multicolumn{7}{c}{(d)\ \ O}\\
 & $\Gamma$ & $X$ & $R$ & $S$ & $T$ & $W$\\
\hline
Band 1\ \ & $\Gamma^+_1$ + $\Gamma^+_2$ + $\Gamma^-_1$ + $\Gamma^-_2$ & $X^+_1$ + $X^+_2$ + $X^-_1$ + $X^-_2$ & $R^+_1$ + $R^-_1$ + $R^+_2$ + $R^-_2$ & $S^+_1$ + $S^-_1$ + $S^+_2$ + $S^-_2$ & 2$T_1$ & 2$W_1$\\
Band 2\ \ & $\Gamma^+_3$ + $\Gamma^+_4$ + $\Gamma^-_3$ + $\Gamma^-_4$ & $X^+_3$ + $X^+_4$ + $X^-_3$ + $X^-_4$ & $R^+_1$ + $R^-_1$ + $R^+_2$ + $R^-_2$ & $S^+_1$ + $S^-_1$ + $S^+_2$ + $S^-_2$ & 2$T_1$ & 2$W_1$\\
\hline\\
\end{tabular}
\ \\
\begin{flushleft}
Notes to Table~\ref{tab:wf74}
\end{flushleft}
\begin{enumerate}
\item The antiferromagnetic structure of LaFeAsO depicted in 
Fig.~\ref{fig:structures} (b) has the space group $Imma$ and the magnetic
group $M = Imma + \{K|\frac{1}{2}\frac{1}{2}0\}Imma$ with $K$ denoting the
operator of time-inversion.   
\item Each row defines one band consisting of four branches, because in each case there
  are four atoms in the unit cell.
\item The representations are given in Table~\ref{tab:rep74}.
\item The bands are determined by Eq.~(23) of Ref.~\protect\cite{josla2cuo4}.
\item Assume a band of the symmetry in any row of theses Tables~\ref{tab:wf74} (a), (b), (c), 
  or (d)   to exist in the band structure of LaFeAsO. Then the Bloch functions of this 
  band can be unitarily transformed into Wannier functions that are
\begin{itemize}
\item as well localized as possible; 
\item centered at the assigned (La, Fe, As, or O) atoms;
\item and symmetry-adapted to the space group $Imma$.
\end{itemize}
\item Eq.~(23) of Ref.~\protect\cite{josla2cuo4} makes sure that the Wannier
  function may be chosen to be symmetry-adapted to the space group $Imma$. In
  addition, there exists a Matrix {\bf N} satisfying both Eqs.~(26) (with
  $\{K|\frac{1}{2}\frac{1}{2}0\})$ and~(32) of Ref.~\protect\cite{josla2cuo4}
  for all the bands listed in this table. Hence, the Wannier functions may be
  chosen symmetry adapted to the magnetic group $M = Imma +
  \{K|\frac{1}{2}\frac{1}{2}0\}Imma$.
\end{enumerate}
\end{table}


\end{document}